\documentclass[twocolumn]{aastex631}

\usepackage{amsmath}
\usepackage{float}
\usepackage{hyperref}
\usepackage{graphicx}
\usepackage{cancel,soul,ulem}
\newcommand{\oiii}{[O\thinspace{\sc iii}]}

\newcommand{\ciii}{[C\thinspace{\sc iii}]}
\newcommand{\civ}{[C\thinspace{\sc iv}]}
\newcommand{\ha}{H$\alpha$}
\newcommand{\hb}{H$\beta$}

\received{July 5, 2022}
\revised{September 22, 2022}
\accepted{October 3, 2022}
\submitjournal{ApJ}

\shorttitle{Triggering of extreme starburst events}
\shortauthors{Laufman et al.}

\begin{document}

\title{On the triggering of extreme starburst events in low-metallicity galaxies: a deep search for companions of Green Peas}

\correspondingauthor{Lauren Laufman}
\email{laufm008@umn.edu}

\author[0000-0002-2121-5137]{Lauren Laufman}
\affiliation{Minnesota Institute for Astrophysics, University of Minnesota, 116 Church St SE, Minneapolis, MN 55455, USA}

\author[0000-0002-9136-8876]{Claudia Scarlata}
\affiliation{Minnesota Institute for Astrophysics, University of Minnesota, 116 Church St SE, Minneapolis, MN 55455, USA}

\author[0000-0001-8587-218X]{Matthew Hayes}
\affiliation{Department of Astronomy and Oskar Klein Centre, Stockholm University, AlbaNova University Centre, 106 91 Stockholm, Sweden}

\author[0000-0003-0605-8732]{Evan Skillman}
\affiliation{Minnesota Institute for Astrophysics, University of Minnesota, 116 Church St SE, Minneapolis, MN 55455, USA}

\begin{abstract}

Green pea galaxies are starbursting, low-mass galaxies that are good analogues to star-forming galaxies in the early universe. We perform a survey of 23 Green Peas using the MUSE Integral Field Unit spectrograph on the VLT to search for companion galaxies. The survey reaches an average point-source depth of $\sim 10^{-18}$ erg cm$^{-2}$ s$^{-1}$ for emission lines. The MUSE field of view allows us to probe a 1$\times$1 arcmin$^2$ field around these galaxies and to search their surroundings for faint companions that could have interacted with them and induced their starburst episodes. We search for companions using a variety of methods including template matching to emission and absorption line spectra. When restricting the search to the same physical area (R = 78 kpc) for all galaxies, we find that the fraction of green pea galaxies with companions is $0.11_{-0.05}^{+0.07}$. We define a control sample of star-forming galaxies with the same stellar masses and redshifts as the green peas, but consistent with the star-formation main sequence. We find that green pea galaxies are as likely to have companions as the control sample; for which the  fraction of objects with companions is $0.08_{-0.03}^{+0.05}$. Given that we do not find statistical evidence for an elevated companion fraction in the green peas in this study, we argue that the ``companions" are likely unrelated to the bursts in these galaxies.

\end{abstract}


\section{Introduction} \label{sec:intro}

Green pea galaxies (hereafter green peas) are a type of compact starbursting galaxy first identified by citizen scientists in the Galaxy Zoo project, that classified galaxies from Sloan Digital Sky Survey (SDSS) images by visual morphologies \citep{lintott_galaxy_2008}. Their name comes from their unresolved shape in SDSS images, and their green color is the result of a strong [OIII]$\lambda 5007$\AA\ emission line. For green peas, [OIII]$\lambda 5007$\AA\ has an unusually large equivalent width (EW) of up to 1000\AA, at least compared to the majority of local galaxies at similar magnitudes. At their redshift of 0.1-0.3, this emission line is shifted into the r-band SDSS filter, which is mapped to green, giving them their visually green appearance in the SDSS images. They typically have low masses (log$(M_*/M_{\odot})\approx 9$), high specific star formation rates (sSFR $\approx$ 10$^{-8}$ yr$^{-1}$), low metallicities ($\sim 20\%$ solar), low reddenings ($E(B-V)\le 0.25$), and are in low density environments \citep{cardamone_galaxy_2009,amorin_oxygen_2010,izotov_green_2011}. Green peas are remarkable as they represent the class of galaxies with the most intense specific star formation in the local universe; they are likely in a short and extreme period of their evolution, given that their extreme sSFR is unsustainable with short gas consumption timescales. Other features of note are: elevated N/O ratios for a given metallicity; models suggest a history dominated by short episodic bursts of star formation \citep{amorin_oxygen_2010,amorin_star_2012}; and green peas tend to be in relatively underdense and isolated regions of space \citep{cardamone_galaxy_2009,brunker_environments_2022}.

The properties of green peas are similar to those measured for very high-redshift galaxies: For example, their specific star formation rates are in line with the high-sSFRs seen at redshift 2 and greater \citep{santini_star_2009,stark_keck_2013,santini_star_2017}, and evidence suggests that high-redshift galaxies beyond $z>7$ have extreme \oiii\ and \hb\ emission lines, with equivalent widths of the same order of magnitude as green peas \citep{roberts-borsani_z_2016}. In addition, the UV spectral properties of green peas and high-redshift star forming galaxies are similar, with Ly$\alpha$, \ciii, and \civ\ commonly observed in both;  \citep{schaerer_strong_2022,ravindranath_semiforbidden_2020} for green peas, \citep{mainali_evidence_2017,stark_spectroscopic_2015,stark_ly_2017,rigby_c_2015} for high-redshift galaxies.
Finally, to strengthen the similarity between green peas and high redshift galaxies, results over the past decades have demonstrated that these galaxies are the only population of galaxies that consistently shows a large fraction of Lyman Continuum (LyC) radiation escaping the interstellar and circumgalactic medium, making them compelling nearby analogs to more distant galaxies to understand the processes regulating the escape of ionizing radiation from galaxies \citep{izotov_eight_2016,izotov_detection_2016,flury_low-redshift_2022-1,flury_low-redshift_2022}.

\subsection{Are the starbursts in green peas triggered by interactions?}

A possible cause of the starbursting behavior of green peas could be that they are externally triggered by interactions with nearby companion galaxies. It has long been known that galaxy interactions can result in episodes of enhanced star-formation: \citet{toomre_galactic_1972} established the now well-known relationship between mergers and disturbed morphological features (e.g., tidal tails), while \citet{larson_star_1978} noted that these disturbed galaxies also tended to have a wider distribution in color, suggesting star formation activity. It has since been strongly established that a majority of local starburst events in galaxies are associated with major merger events \citep{kennicutt_induced_1984, joseph_recent_1985, armus_multicolor_1987, sanders_luminous_1996, farrah_hstwfpc2_2001, arribas_optical_2004}. Although merger-induced star formation is expected to be short-lived and contribute only a small amount to the total star formation in the local universe \citep{robaina_less_2009,jogee_history_2009}, it is a key mechanism for explaining extreme starburst events. A recent study \citep{cibinel_early-_2019} found that, at redshifts between $z \sim 0.2$ and $z \sim 2$, virtually all starbursting galaxies (those well above the star-forming main sequence (MS) with $\Delta_{MS} = log(SFR) - log(SFR_{MS}) > 0.6$) show signs of an ongoing major merger. By comparison, only 5-15\% of main sequence galaxies show major merger activity over this redshift range. 
It has been well-established that the major merger rate increases with redshift  \citep[e.g.][]{duncan_observational_2019,lotz_major_2011}, which may help explain the decline in highly star forming galaxies over this time frame.
\citet{ellison_galaxy_2008} placed a distance limit on this effect by noting that star formation was most enhanced for pairs closer than 30 $h^{-1}$ kpc, while it was not significantly enhanced for galaxies farther apart from this. This effect was most pronounced for pairs of galaxies with similar masses (within a factor of 2).

While the aforementioned studies mostly discuss mergers of larger and redder galaxies, it has also been shown that mergers can induce star formation in dwarf galaxies. For example, \citet{weisz_recent_2008} studied dwarf irregular galaxies in the M81 Group, and found in DDO 53 that the SFR was enhanced about 1 Gyr ago. A dwarf-dwarf merger scenario was proposed and they claimed it could explain the starburst and the peculiarities observed in its appearance. In a multiwavelength study of DDO 53, in particular examining the anomalous HI distribution and kinematics, \citet{egorov_star_2021} came to a similar conclusion as \citet{weisz_recent_2008}. Similarly, \citet{zhang_blue_2020} performed a multiwavelength study of blue compact dwarf galaxy VCC 848, and identified a clear link between a dwarf-dwarf merger event and enhanced star formation. Another example is the blue compact dwarf galaxy Haro 14. \citet{cairos_mapping_2010} used the Multi Unit Spectroscopic Explorer (MUSE) at the Very Large Telescope to study Haro 14, and found that the lopsidedness and the presence of multiple stellar clusters are consistent with a merger or interaction scenario. Among the closest analogues to green peas are luminous compact galaxies (LCGs), which share similar masses, luminosities, and metallicity to the green peas \citep{cardamone_galaxy_2009,izotov_green_2011}. LCGs are difficult to study due to their rarity and compact morphology, but an HST study of $\sim$40 LCGs found that about 36\% of these had morphologies consistent with ongoing mergers \citep{rawat_unravelling_2007}.

Given the extreme star-forming nature of green peas, it is worth placing them in context compared to other local, compact, star-forming galaxies. In particular, green peas have often been compared to blue compact dwarf (BCD) galaxies. \citet{gil_de_paz_palomarcampanas_2003} defined BCDs as having the following properties: First, they are blue; they are required to have $\mu_{B,peak} - \mu_{R,peak} \lesssim 1$. Second, they are compact; they have to have a relatively small high-surface-brightness component, $\mu_{B,peak} < $ 22 mag arcsec$^{-2}$. Third, they have to be dwarfs, limiting their K-band luminosity to $M_{K,Vega} > -21$ mag. Green peas show significant differences from BCDs, as discussed in \citet{cardamone_galaxy_2009}. While they are similar in morphology, environment, and size, green peas have higher stellar masses by an order of magnitude than BCDs, and are also more luminous than BCDs. Green peas are more closely related to luminous blue compact galaxies (LCGs). However, given that LCGs are rarer and typically more distant than BCDs, there is value in examining the known behavior of BCDs as an indirect way of assessing green peas.

BCDs have been investigated in the past to see if galaxy interactions are a potential cause for BCDs’ intense star formation. \cite{taylor_h_1997} conducted a VLA HI study of 21 H\thinspace{\sc ii} (BCD) galaxies and 17 low surface brightness (LSB) dwarf galaxies,  finding that  BCDs  had companions more than twice as often as the LSB dwarfs. Additionally, a multiwavelength study by \citet{lopez-sanchez_environment_2010} of nearby BCD galaxies found that all showed evident interaction features in their neutral gas component, and suggested that interactions are the main trigger mechanism of starbursting behavior in BCD galaxies. On the other hand, \citet{telles_environment_1995} conducted an optical search for companions in the neighborhood of 51 BCD galaxies to check the tidal origin of starbursts. They found that the isolated BCDs were of high luminosity and disturbed morphology, while the BCDs with neighbors were low luminosity and regular morphology. They also found that the metal abundance and equivalent width of the emission lines in their BCD sample galaxies did not depend on having a companion. While the relationship between interactions and star formation in BCDs is not fully established, the possibility of a connection between the two suggests that it may be worthwhile to examine green peas for this effect.

In this paper, we investigate a sample of 23 green pea galaxies and search for evidence of nearby companions that may have influenced their star formation history. Even though \citet{cardamone_galaxy_2009} studied the environment of these galaxies, their results are based on the shallow SDSS spectroscopic survey, with a continuum pre-selection of targets. Here, we present a new survey that uses large-field integral-field spectroscopy to revisit the link between interactions and starbursts in green peas. In Section \ref{sec:data}, we discuss data acquisition and reduction. In Section \ref{sec:findingcomp}, we discuss methods of searching for companions. Section \ref{sec:comptocontrol} compares the fraction of green peas with a companion to a control sample of galaxies. Results and discussion are in Section \ref{sec:resdis}, and conclusions in Section \ref{sec:summ}.

\section{Data}
\label{sec:data}

\subsection{Data Acquisition}

We selected a subsample of 24 green peas from the original catalog defined by \citet{cardamone_galaxy_2009}. The green peas were selected such that they would be visible from Cerro Paranal, the location of the Very Large Telescope (VLT), during the period that the observations were taken (from October 2018 to March 2019). The data were taken using the Multi Unit Spectroscopic Explorer (MUSE) \citep{bacon_muse_2010}, an Integral Field Unit (IFU) spectrograph, on the VLT Unit Telescope 4 at the European Southern Observatory. The field of view covers 1$\times$1 arcmin$^2$ and the spectral range is 0.475-0.91 $\mu$m. Large field-of-view IFUs, such as MUSE, are particularly well-suited to identifying nearby companions to the green peas because not only do they take an image, but each pixel is a spaxel, an entire spectrum (with the aforementioned spectral range). The pixel scale of the instrument is 0.2x0.2 arcsec$^2$ per square pixel. The spectral resolution is 1750 at 465 nm and 3750 at 930 nm. The resulting data product is a data cube covering a 1$\times$1 arcmin$^2$ field around each green pea. This corresponds to a projected physical distance of 159 - 271 kpc at the distance of our targets (redshifts 0.15 - 0.3).

The total integration time for each green pea was at least 2800 seconds, split into 700 seconds per exposure for four exposures. The required conditions for observations to take place were as follows: seeing $<$ 2\arcsec, sky transparency: thin, airmass $<$ 2, and moon distance $>$ 30 degrees. Most of the observations were observed with seeing better than 1.5, see Table \ref{tab:info} for details. The observations were taken in service mode. If the seeing conditions were not fulfilled, then the four exposures would be taken again on another night; this led to some green peas having up to seven exposures in total, which were all used during the data reduction process. The specific exposure lengths and measured seeing for the final reduced cubes are in Table \ref{tab:info}.

\begin{deluxetable*}{lccccccc}
\tablecaption{The green pea galaxies Sample\label{tab:info}}
\tabletypesize{\scriptsize}
\tablehead{
\colhead{ID}	& \colhead{RA}	& \colhead{Dec}	& \colhead{log($M_*/M_{\odot}$)\tablenotemark{a}}	& \colhead{z\tablenotemark{b}}	& \colhead{Detection Limit (at H$\alpha$)}		&  \colhead{Seeing\tablenotemark{c}}	& \colhead{Exp. Time}\\
\colhead{}		& \colhead{}	& \colhead{}	& \colhead{(Izotov)} 				& \colhead{} 			& \colhead{erg cm$^{-2}$ s$^{-1}$}		&  \colhead{arcsec} 			& \colhead{ s}
}
\startdata
J094458.22-004545.4 & 146.242618 & -0.762639 & 9.76*    & 0.3002 & 7.96E-19 & 0.99   & 4900     \\
J130211.15-000516.4 & 195.54646  & -0.087897 & 9.02     & 0.2255 & \nodata        & \nodata      & 2800     \\
J232539.23+004507.2 & 351.413453 & 0.752012  & 9.41     & 0.2770 & 1.19E-18 & 1.02   & 2800     \\
J032244.89+004442.3 & 50.687082  & 0.745111  & 9.89*    & 0.3043 & 7.93E-19 & 0.90   & 2800     \\
J012910.15+145934.6 & 22.292299  & 14.992956 & 9.46*    & 0.2800 & 2.86E-18 & 0.87   & 2800     \\
J030321.41-075923.2 & 45.839226  & -7.989791 & 9.15     & 0.1650 & 6.94E-19 & 0.79   & 2100     \\
J032613.62-063512.5 & 51.556792  & -6.586816 & 9.48*    & 0.1621 & 4.09E-18 & 1.45   & 2800     \\
J033947.79-072541.2 & 54.949128  & -7.428132 & 9.7      & 0.2608 & 8.71E-19 & 0.89   & 4900     \\
J105716.72+023207.0 & 164.3197   & 2.535293  & 9.90*    & 0.3028 & 1.97E-18 & 1.02   & 2800     \\
J124423.37+021540.4 & 191.097382 & 2.261231  & 9.65     & 0.2395 & 1.54E-18 & 1.06   & 2800     \\
J084216.95+033806.6 & 130.57063  & 3.635203  & 9.38     & 0.2194 & 1.79E-18 & 0.91   & 4200     \\
J154709.10+033614.0 & 236.787938 & 3.603914  & 8.79     & 0.2314 & 1.96E-18 & 1.36   & 2800     \\
J223735.05+133647.0 & 339.396081 & 13.613062 & 9.45     & 0.2936 & 1.28E-18 & 1.36   & 2800     \\
J133711.88-022605.4 & 204.299529 & -2.434842 & 9.19     & 0.2737 & 1.34E-18 & 1.34   & 2800     \\
J144231.37-020952.0 & 220.630713 & -2.164466 & 8.65     & 0.2938 & 1.47E-18 & 1.31   & 2800     \\
J103138.93+071556.5 & 157.912214 & 7.265701  & 8.75     & 0.2525 & 9.03E-19 & 0.77   & 2800     \\
J124834.63+123402.9 & 192.14431  & 12.56748  & 9.05     & 0.2634 & 1.73E-18 & 0.84   & 2800     \\
J160436.66+081959.1 & 241.152768 & 8.333082  & 8.48     & 0.3123 & 2.42E-18 & 1.71   & 2800     \\
J155925.97+084119.1 & 239.858241 & 8.688655  & 8.97     & 0.2970 & 1.16E-18 & 0.92   & 2800     \\
J101157.08+130822.0 & 152.98785  & 13.139471 & 8.31     & 0.1439 & 5.86E-19 & 0.70   & 2800     \\
J163719.30+143904.9 & 249.330431 & 14.651378 & 9.41     & 0.2923 & 1.64E-18 & 1.60   & 2800     \\
J092532.36+140313.1 & 141.384863 & 14.053623 & 8.46     & 0.3013 & 8.23E-19 & 0.76   & 2800     \\
J161306.31+092949.1 & 243.276317 & 9.49699   & 9.65*    & 0.2993 & 9.39E-19 & 1.27   & 2800     \\
J080518.06+092533.3 & 121.325174 & 9.425978  & 9.36     & 0.3304 & 1.44E-18 & 1.15   & 2800     \\
\enddata
\tablenotetext{a}{Reproduced from \citet{izotov_green_2011}, unless marked with a ``*", in which case \citet{izotov_green_2011} did not calculate a mass for that galaxy, then the \citet{cardamone_galaxy_2009} value was used instead. \citet{izotov_green_2011} recalculated the masses of the green pea galaxies and got systematically lower values. Their values are lower because in fitting the SED, they subtracted the contribution from gaseous continuum emission. No errors were listed in the original studies.}
\tablenotetext{b}{Reproduced from \citet{cardamone_galaxy_2009}}
\tablenotetext{c}{The seeing is calculated for the 6500-7300\AA\ wavelength range.}
\tablenotetext{d}{We were unable to reduce the data for J130211.}
\end{deluxetable*}

\begin{deluxetable*}{lcccccccc}
\tablecaption{Green pea galaxies with companions and their properties\label{tab:info2}}
\tabletypesize{\scriptsize}
\tablehead{
\colhead{ID}	&\colhead{Continuum Mag}				& \colhead{H$\alpha$ Flux} & \colhead{H$\alpha$ EW}	& \colhead{Distance}		& \colhead{Distance}	& \colhead{Companion H$\alpha$ Flux} &   \colhead{Companion \ha\ EW}&   \colhead{Companion}\\
\colhead{}	& \colhead{(gemini\_dssi\_red)\tablenotemark{a}}	& \colhead{(erg cm$^{-2}$ s$^{-1}$})	& \colhead{(erg cm$^{-2}$ s$^{-1}$)} & \colhead{(arcsec)}		& \colhead{(kpc)}	& \colhead{(erg cm$^{-2}$ s$^{-1}$)} & \colhead{(erg cm$^{-2}$ s$^{-1}$)}   & \colhead{\hb\ EW (\AA)}
}
\startdata
J094458.22-004545.4 & 20.21    & $9.41\pm0.01 \times 10^{-16}$ & $249.9\pm0.2$  & 26.78   & 120.38   & $7.34\pm0.03 \times 10^{-17}$ & $31.7\pm0.15$     &  $8.2\pm0.17$\\
J124834.63+123402.9 & 20.29    & $2.19\pm0.00 \times 10^{-15}$ & $589.9\pm0.8$  & 25.14   & 102.99   & $8.33\pm0.08 \times 10^{-17}$ & $28.4\pm0.29$     & $11.4\pm0.17$\\
J163719.30+143904.9 & 21.06    & $6.40\pm0.01 \times 10^{-16}$ & $348.1\pm0.5$  & 13.19   & 58.20    & $3.38\pm0.05 \times 10^{-17}$ & $37.2\pm0.61$     & $24.2\pm1.79$\\
J161306.31+092949.1 & 20.04    & $8.93\pm0.01 \times 10^{-16}$ & $184.8\pm0.2$  & 13.64   & 61.18    & $6.27\pm0.04 \times 10^{-17}$ & $160.6\pm1.01$     & $46.4\pm0.96$\\
\enddata
\tablenotetext{a}{Filter chosen to only capture flux in the continuum of the green peas.}
\end{deluxetable*}

\subsection{Data Reduction}

Data were reduced using the MUSE Data Reduction Software Pipeline EsoRex version 2.8.1 \citep{weilbacher_data_2020}.

We use the MUSE pipeline for the following steps: bias, dark correction and flat field corrections, wavelength calibration, illumination correction, flux calibration, and combining exposures. The individual steps of the MUSE pipeline are relatively standard for data reduction. The bias is subtracted, the darks are subtracted (if darks were taken on a given night), the data are divided by the lamp flat field, and then a polynomial wavelength solution is fit to the calibration arc lamp emission lines. For all of these, the exposures are combined with the combination parameter set to sigclip (sigma clipping), at a value of 15. The line spread function (LSF) and instrument geometry recipes are skipped, as their only purpose is to generate calibration files, and we instead used the provided calibration files included in the pipeline installation. Generating the LSF profile is computationally intensive, and the MUSE Pipeline User Manual \citep{weilbacher_data_2020} explains that the parameters of the LSF are stable so it does not need to be frequently updated. The twilight sky flats are combined and correction factors are computed. This step creates a 3D illumination correction, processes the cube to make a mask of the illuminated area from the white light, then smooths it, fits with a polynomial, and collapses the smooth cube. The standard star frames are then used to create a flux response curve. In the post processing step, the pixel tables of each IFU are merged into a single pixel table for each exposure, and converted back into a fits data cube. Finally, the individual exposures are aligned with each other and then combined using drizzle resampling into the final output product.

This process produces a data cube, along with a white light image. We excluded one green pea from the analysis, leaving 23 instead of 24 green peas to be analyzed. The excluded green pea, J130211.15-000516.4, resulted in errors when attempting to process through the pipeline, despite numerous attempts to resolve the errors. There were also issues with a few green peas during the automatic alignment process of the pipeline. For those green peas we manually input the coordinates of reference stars for the alignment, and these coordinates were obtained using Source Extractor \citep{bertin_sextractor_1996}.

The seeing was calculated by fitting a 2D Gaussian to a compact source within each green pea field. The resulting Gaussian FWHM are reported in Table \ref{tab:info}. The image used for calculating the seeing was the white light image of the final data cube produced by the pipeline.

To determine the line flux limit in a datacube, we estimate the 5$\sigma$ emission line flux limit within a typical aperture size. We make pseudo emission line H$\alpha$ images (described in Section~\ref{sec:resid}), and then we use them to examine galaxy flux within an aperture with diameter equal to the FWHM of the point spread function. To determine the noise, we start by selecting a background region consisting of all pixels in the segmentation map that are not associated with a source. The segmentation map is a mask that has positive values where there is an object and is zero everywhere else, and was generated using Source Extractor. We then examine these pixels in the pseudo emission line H$\alpha$ images. We estimate the mean background $\mu_{bg}$ to be the mean pixel flux in this background region, and estimate the standard deviation of noise per pixel $\sigma_{bg}$ to be the standard deviation in this region. Finally, we estimate the background within a galaxy's aperture to be $\mu_{ap} = N_{ap} \mu_{bg}$, where $N_{ap}$ is the number of pixels within the aperture. Its standard deviation is $\sigma_{ap}$ = $\sqrt{N_{ap}} \sigma_{bg}$. The 5$\sigma$ line flux limits for each green pea field are reported in Table \ref{tab:info}. The average line luminosity limit for H$\alpha$ is approximately $10^{38.5}$ erg s$^{-1}$, which is roughly 1000 times less luminous than the typical green pea targets themselves. This corresponds to a star-formation-rate of 0.003 M$_{\odot}$ yr$^{-1}$, assuming the conversion of \cite{bell_comparison_2001}.

\section{The Search for Companions}
\label{sec:findingcomp}

To search for companions, we first must define what makes something a nearby companion. We define a nearby companion as within the field of view ($\sim$0.5 arcminutes on the sky) and less than 300 km s$^{-1}$ radial velocity separation. The 0.5 arcminutes is half of the MUSE field of view (where the green peas are centered in a $1\times1$ arcminute box). The 300 km s$^{-1}$ is chosen so that we avoid including objects that are at significantly different redshifts, and only select those that could have reasonably been within range of gravitational interaction at some point in their recent history. \citet{ventou_new_2019} recommend 300 to 500 km s$^{-1}$, similar to \citet{ellison_galaxy_2008} and \citet{patton_new_2000}'s choice of 500 km s$^{-1}$ for defining a close companion. The latter two chose those values as a compromise between higher detection rates and contamination. The companion fractions do not differ if 500 km s$^{-1}$ is used instead of 300 km s$^{-1}$.

\subsection{\texorpdfstring{H$\alpha$ and [O\thinspace{\sc iii}] Emission Narrowband Images}{Ha and [OIII] Emission Narrowband Images} }
\label{sec:resid}

We create \ha\ and \oiii\ pseudo emission line maps to identify companions at the same redshift as the green pea, that were also emitting H$\alpha$ and/or \oiii. While green peas often show an \oiii/\ha$>1$ \citep{cardamone_galaxy_2009}, it is not guaranteed that the companions have such high excitation spectra. In emission line galaxies, H$\alpha$ is often a distinctive and easily identifiable and confirmable emission line. It is often the strongest among the optical lines, it is the least sensitive to dust attenuation, and it is insensitive to metallicity. Accordingly, in a search for emission line companions, we started by examining nearby objects with an H$\alpha$ emission line. 

We created a pseudo narrow-band image around the H$\alpha$ wavelength at the redshift of each green pea. We consider a width of the narrow-band image of $\pm$300~km~s$^{-1}$ from the center of the H$\alpha$ emission line. We also created a pseudo narrow-band image around the continuum, which was located at $\pm$2000~km~s$^{-1}$ from the center of the H$\alpha$ emission line. The continuum was chosen to be safely outside of the width of the hydrogen line, and to exclude [N\thinspace{\sc ii}] emission lines. The width of this narrow-band continuum image is approximately 500~km~s$^{-1}$, and this value was chosen get a large signal to noise ratio in the continuum.

\begin{figure*}[ht]
\gridline{\fig{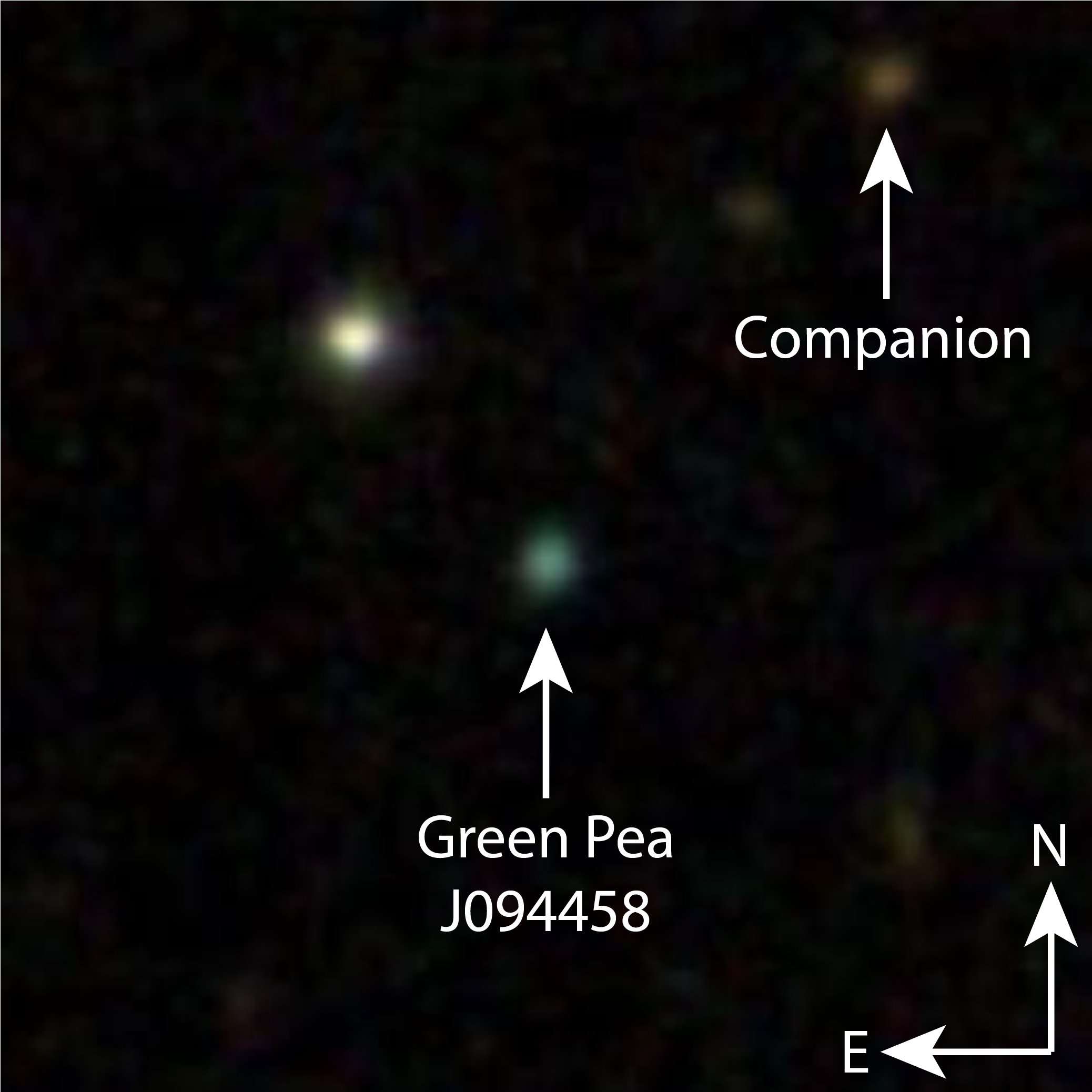}{0.2\textwidth}{(a)}
          \fig{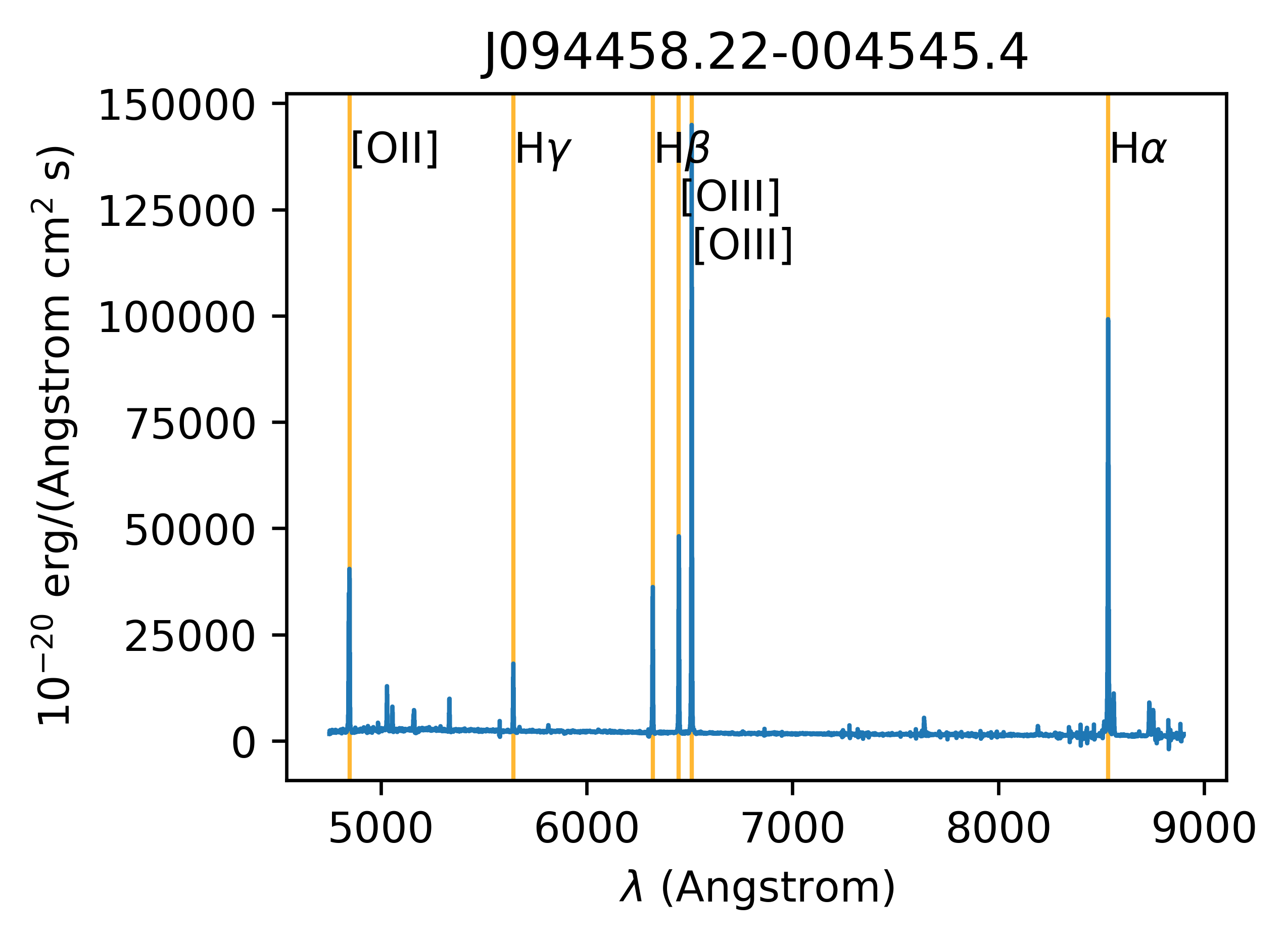}{0.3\textwidth}{(b)}
          \fig{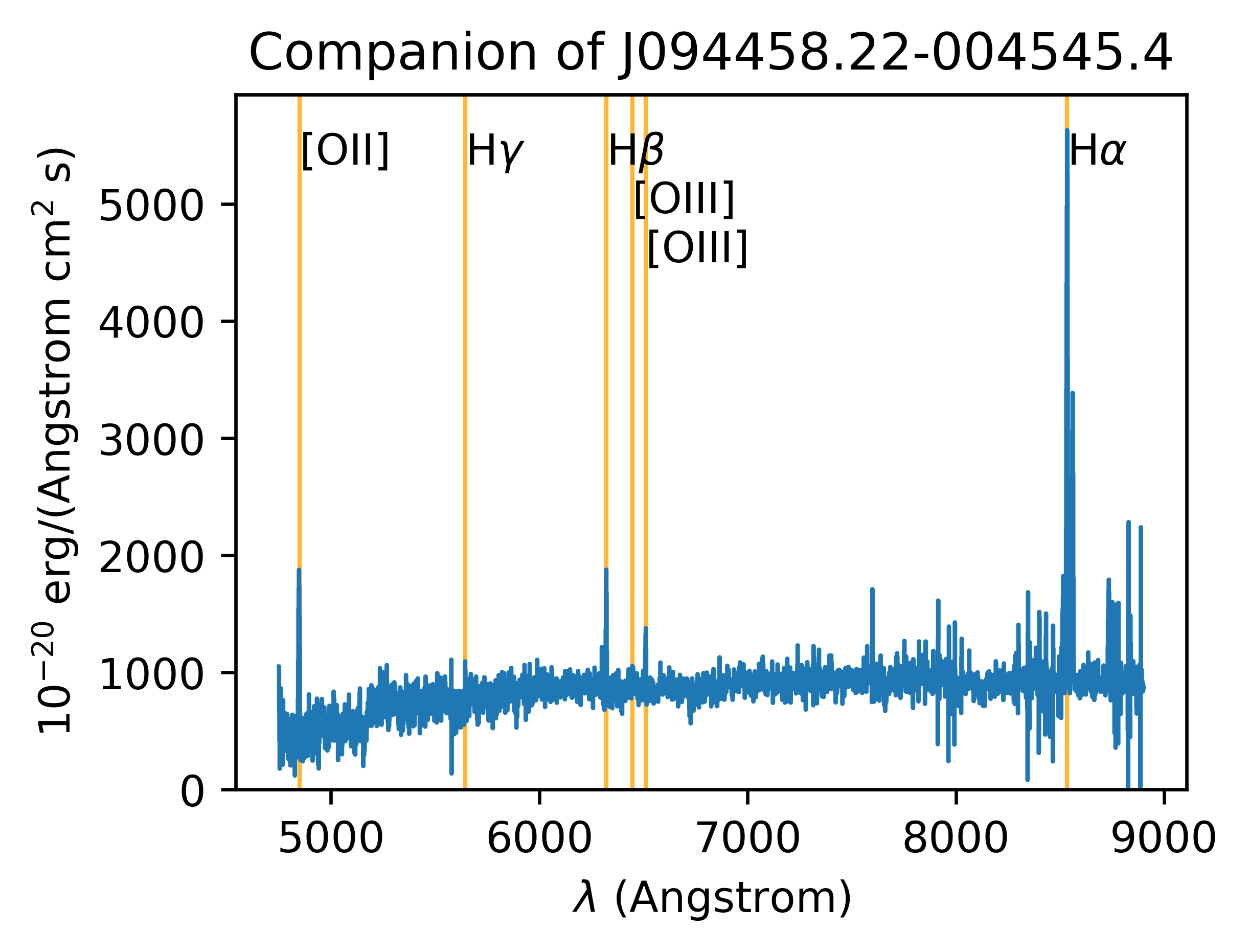}{0.3\textwidth}{(c)}
          }
\gridline{\fig{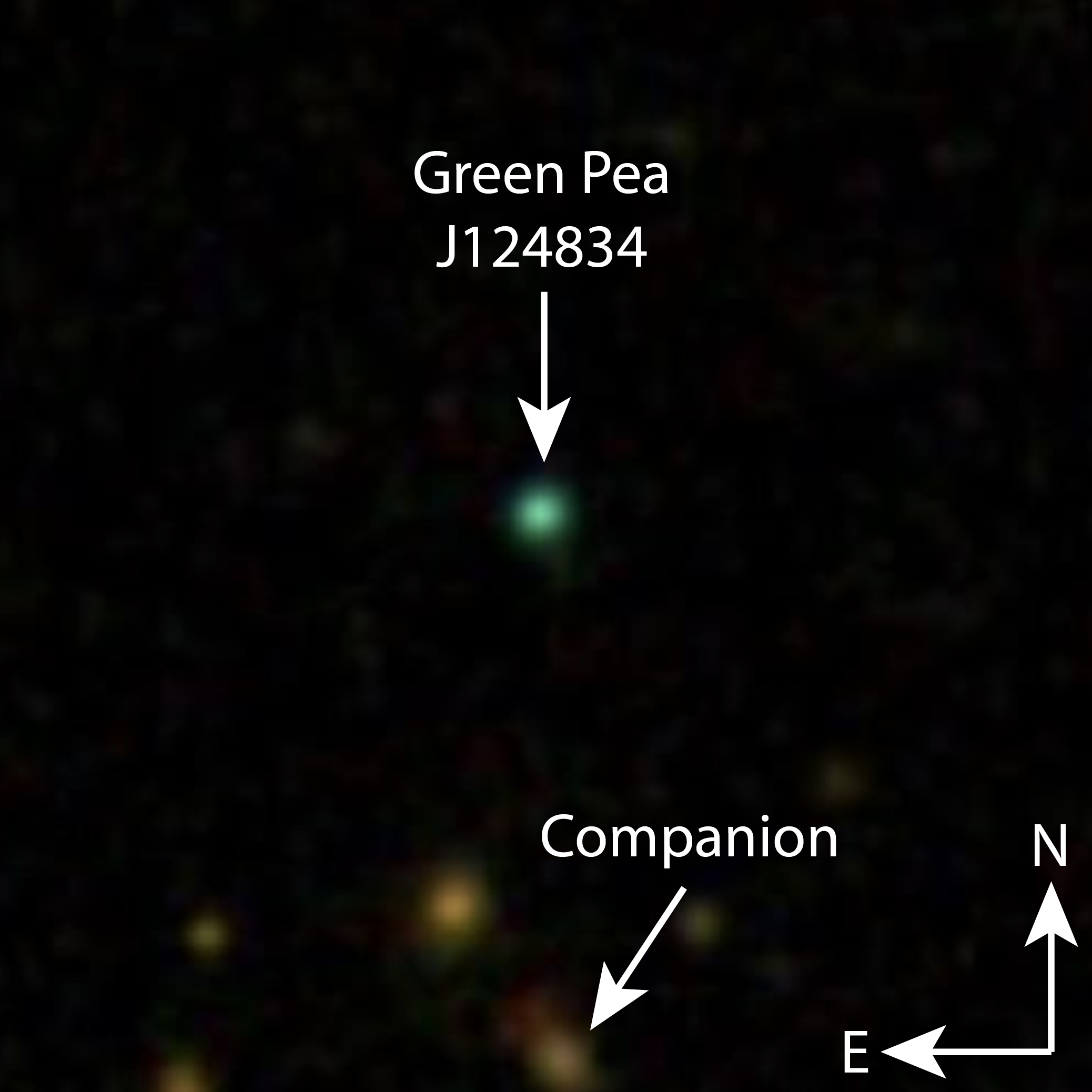}{0.2\textwidth}{(d)}
          \fig{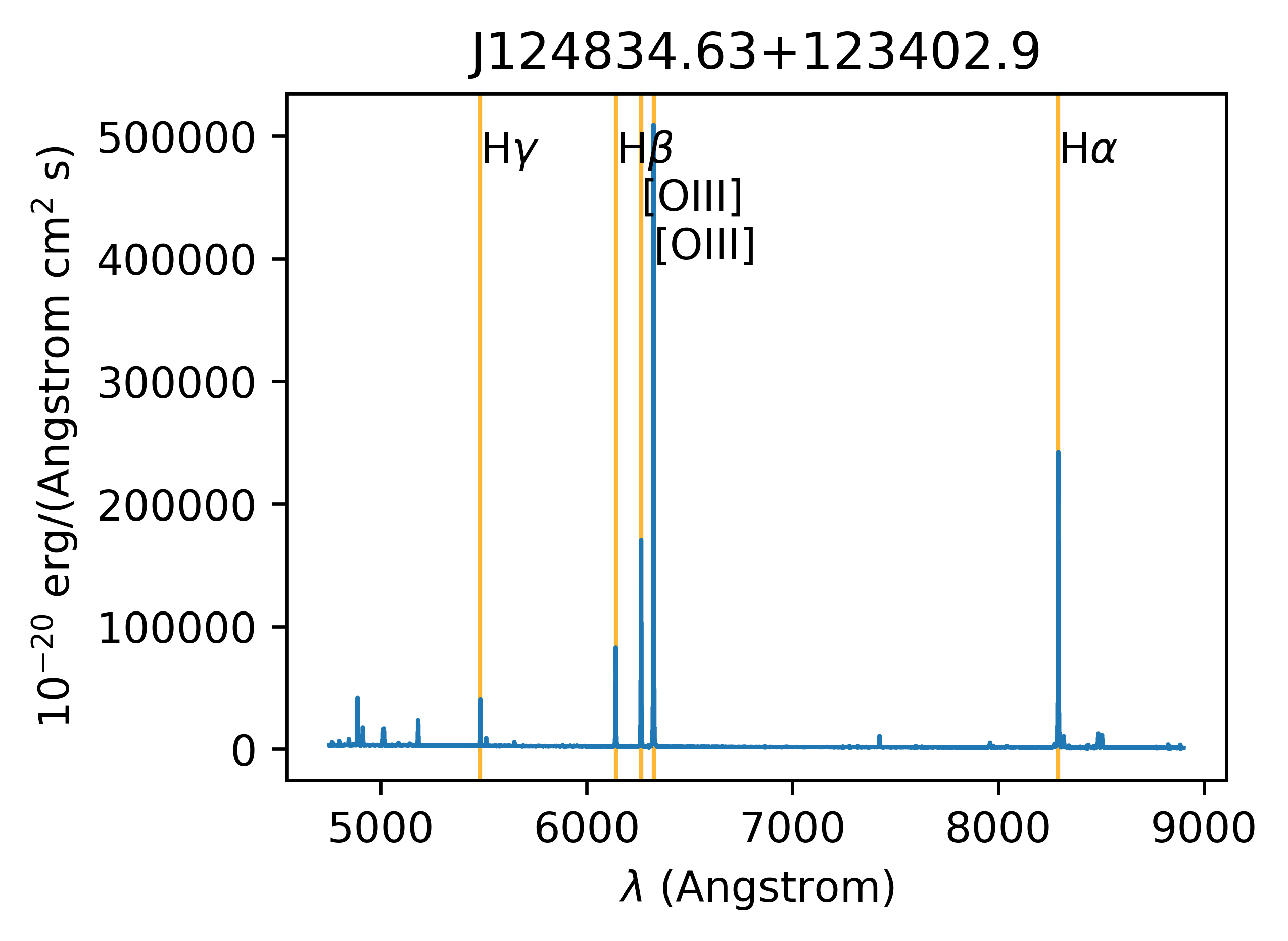}{0.3\textwidth}{(e)}
          \fig{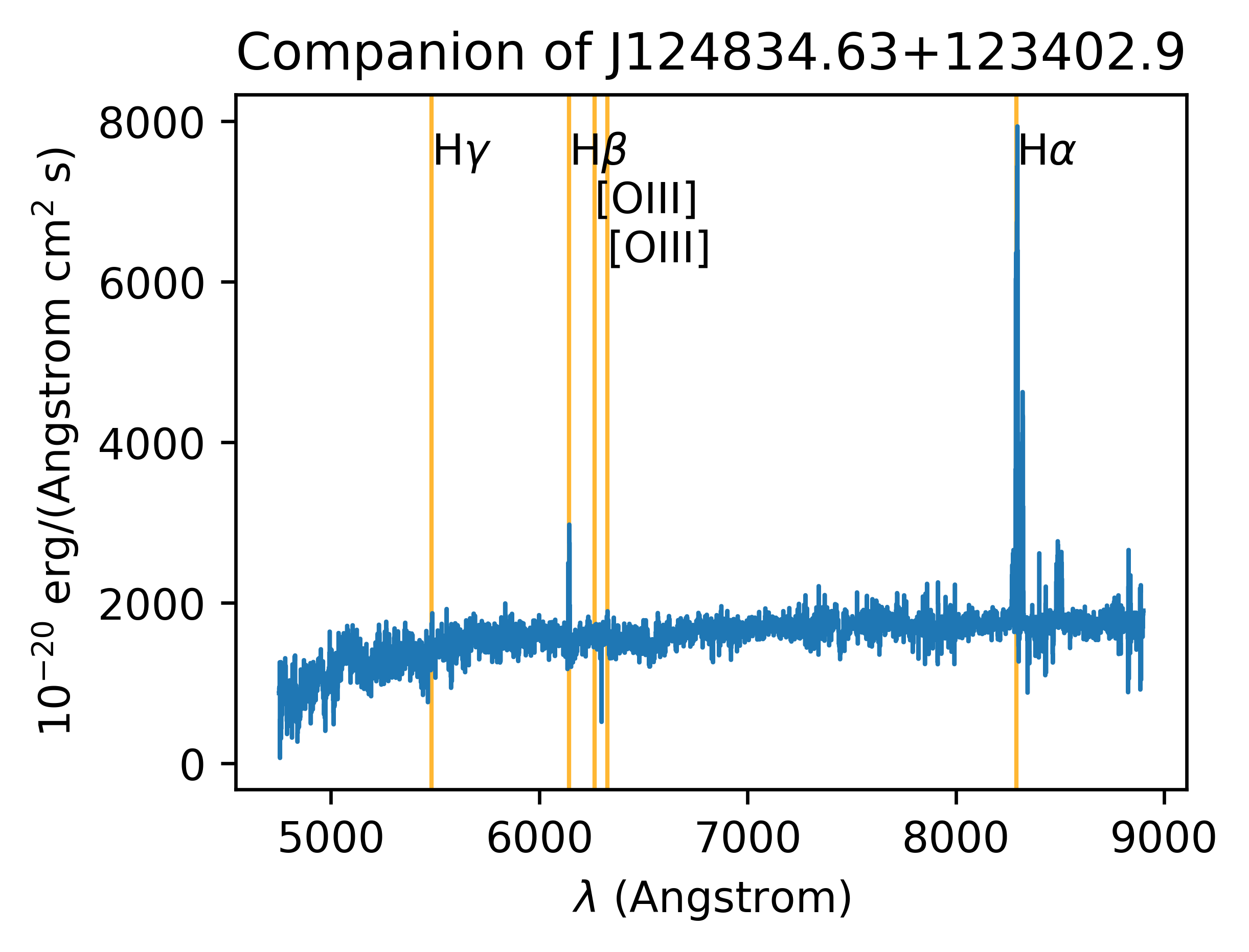}{0.3\textwidth}{(f)}
          }
\gridline{\fig{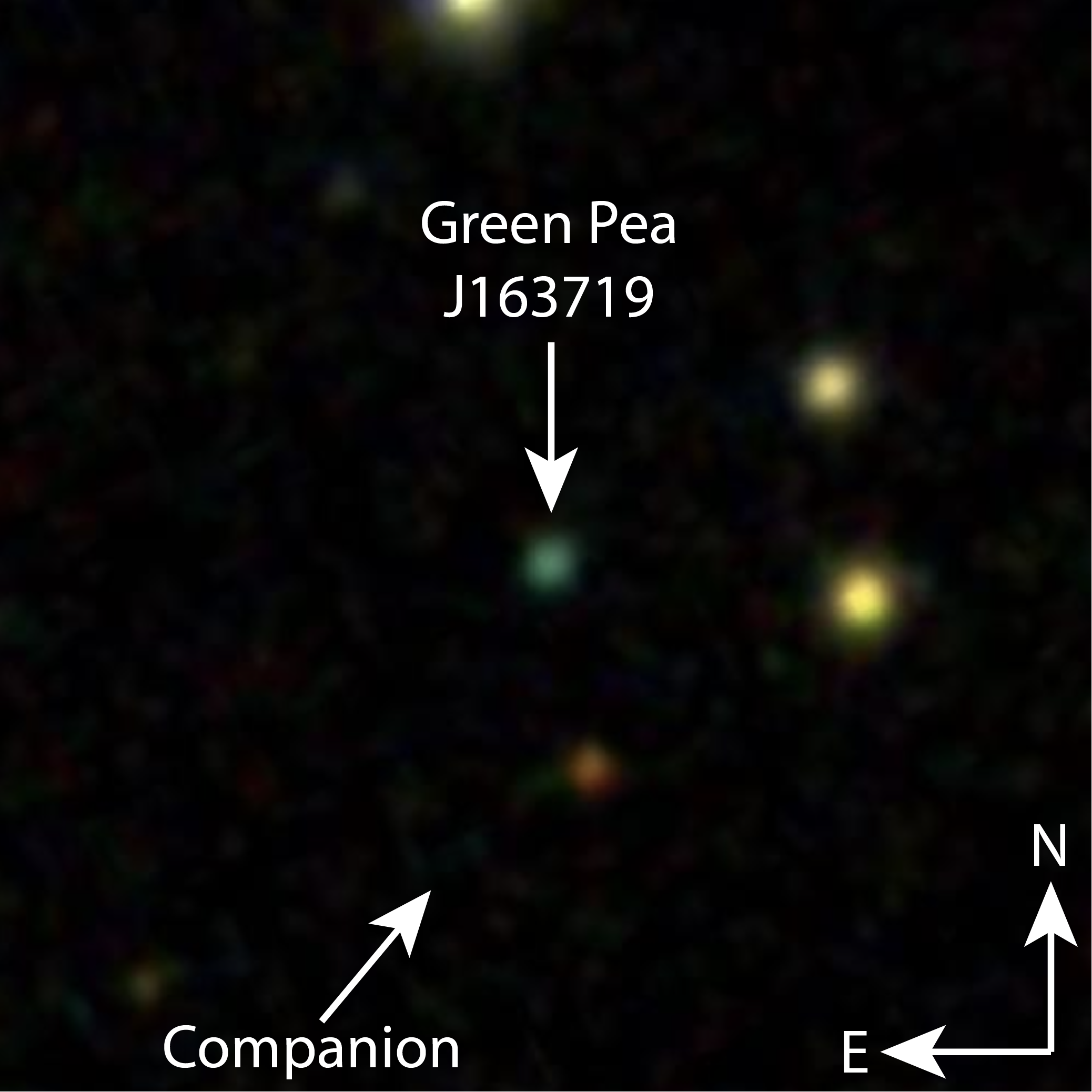}{0.2\textwidth}{(g)}
          \fig{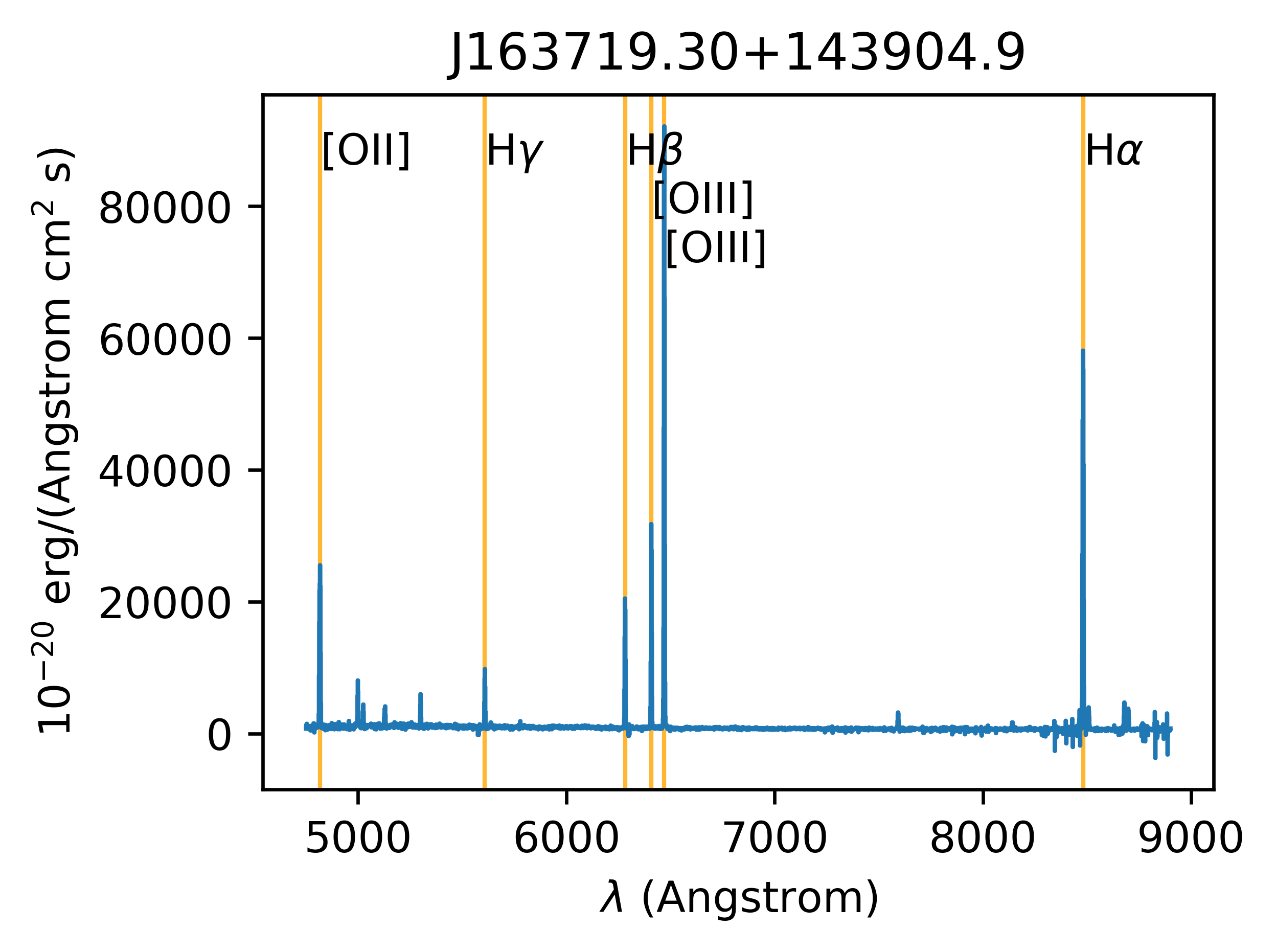}{0.3\textwidth}{(h)}
          \fig{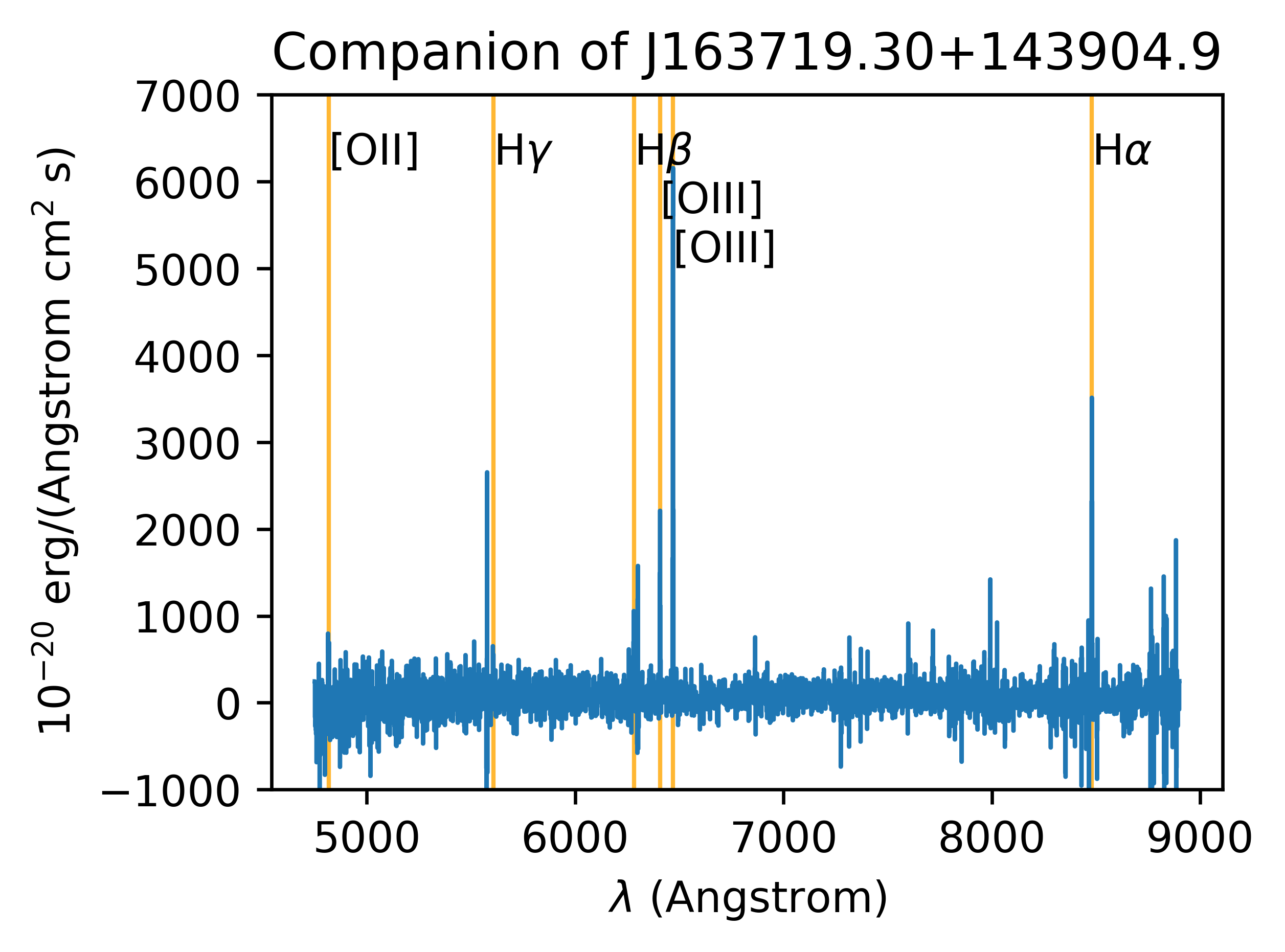}{0.3\textwidth}{(i)}
          }
\gridline{\fig{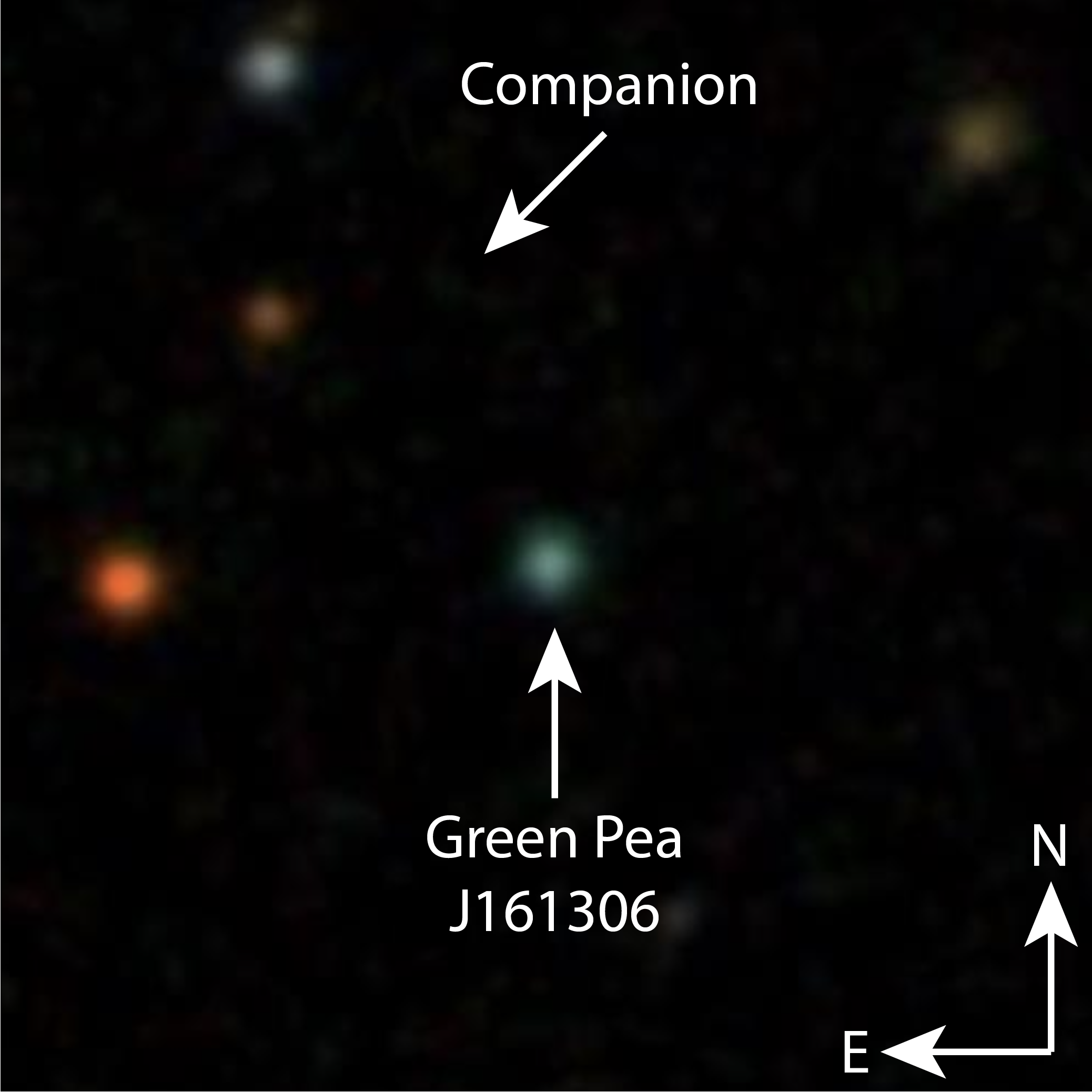}{0.2\textwidth}{j)}
          \fig{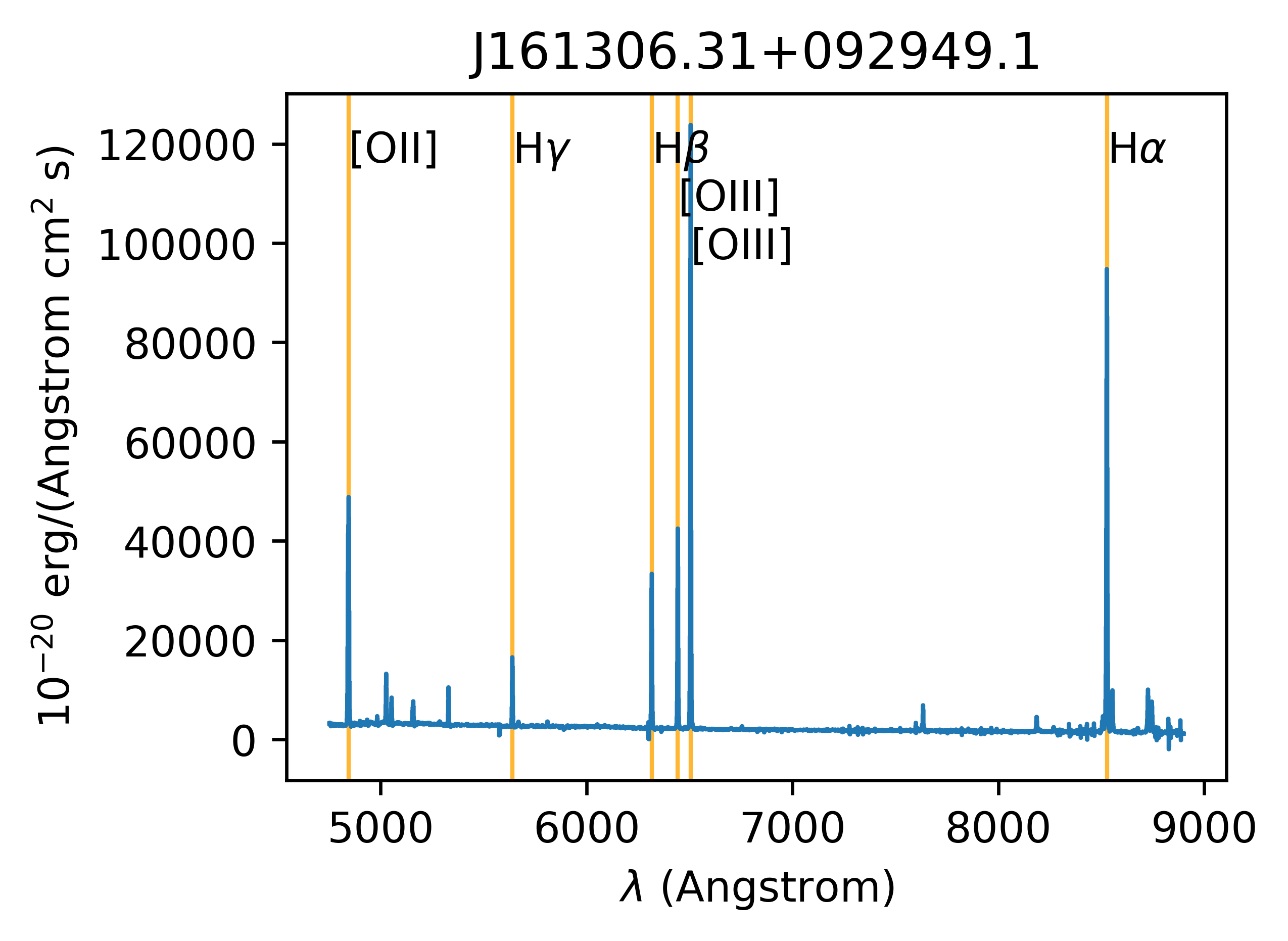}{0.3\textwidth}{(k)}
          \fig{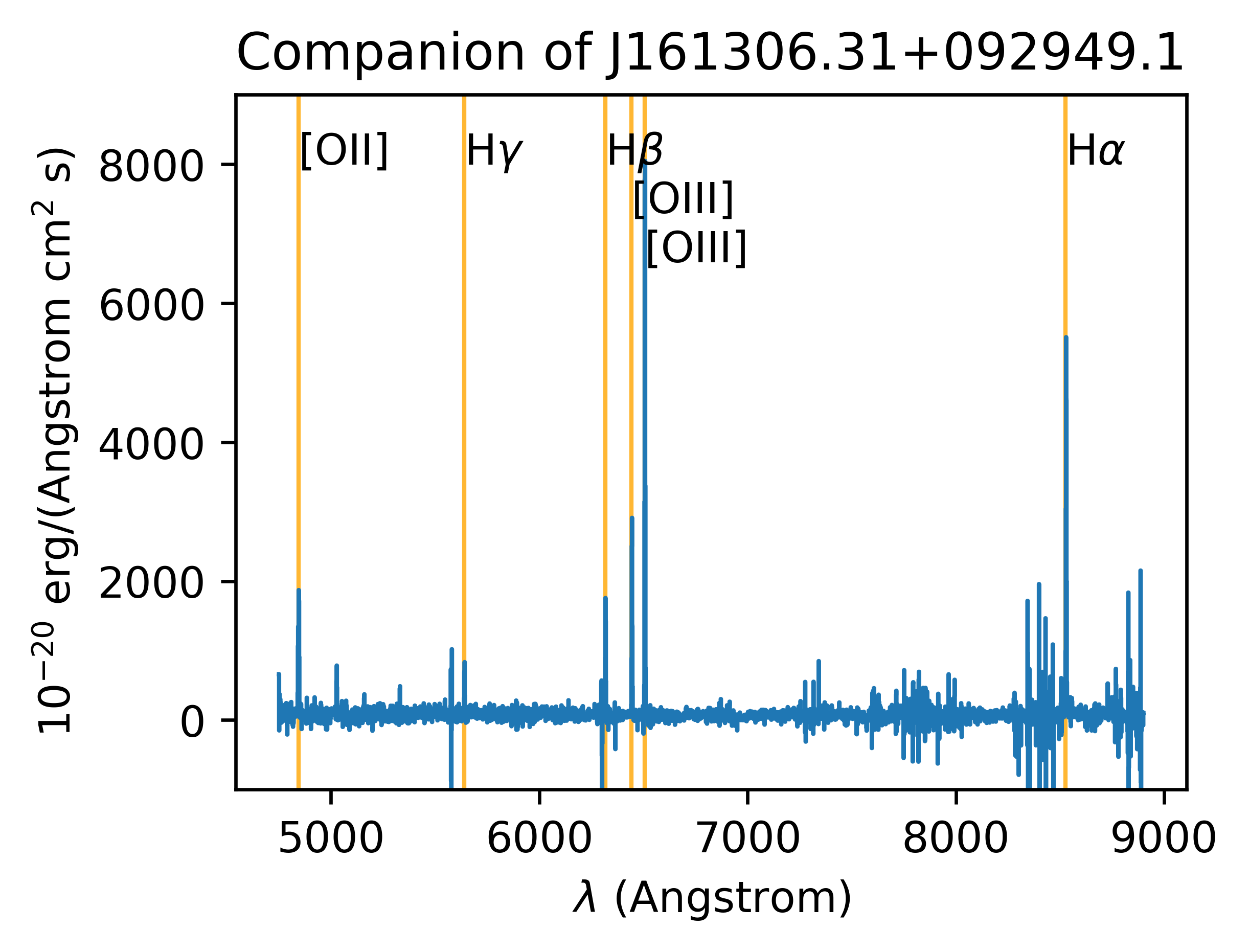}{0.3\textwidth}{(l)}
          }
\caption{Left column: SDSS images of the green peas (J094458, J124834, J163719, J161306) and their companions. In the SDSS images, two of the four companions are not visible. Middle column: Spectra of the aforementioned green peas. Right column: Spectra of the companions of the aforementioned galaxies. The plotted wavelengths are in the observed frame. Companion galaxies have fainter continua compared to the main green pea galaxy, and residual flux from the sky subtraction process is more prominent in their spectra. The main emission line features detected in each spectrum are labeled.
\label{fig:spectra}}
\end{figure*}

\begin{figure*}
\plotone{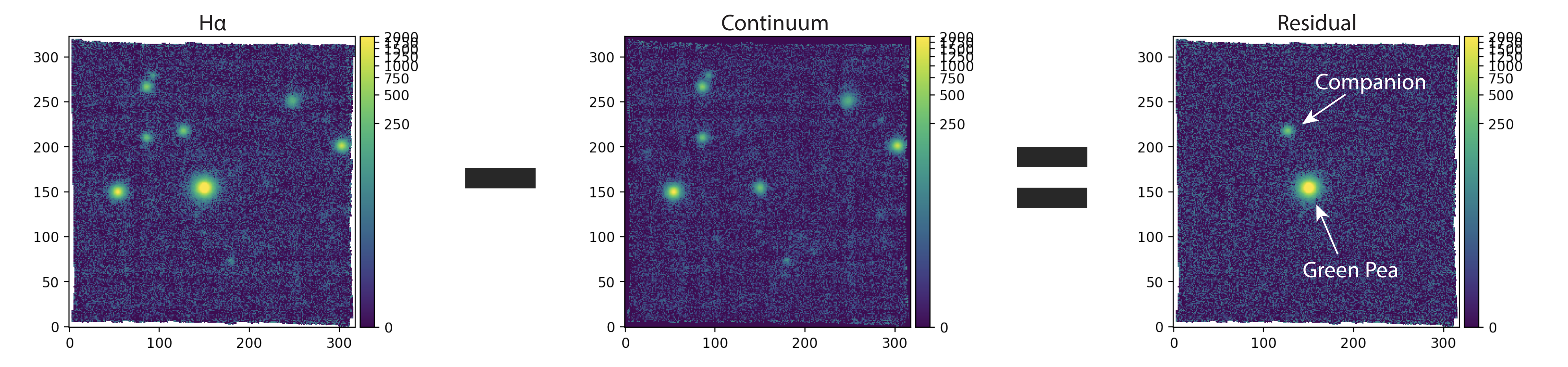}    
\caption{Using the method of subtracting a continuum image from a narrowband image to capture everything emitting specifically at that wavelength. The left image displays the pseudo-narrowband H$\alpha$ image for J161306, the middle image contains the pseudo-narrowband continuum, and the right image is the difference between the two. The green pea and its companion are clearly visible. The x and y axes are in pixels.
\label{fig:residual}}
\end{figure*}

We then subtracted the continuum image from the H$\alpha$ image. This process resulted in an emission line only map, where objects with H$\alpha$ in emission and at the same redshift of the green pea can be identified. Objects at different redshifts and without an emission line do not appear in the line-only image. We then inspected this map for any other objects besides the green pea, and if any were apparent, we checked their spectra for characteristic spectral features of a star forming companion. The features that we looked for were strong \oiii\ and H$\alpha$ emission. An example of the narrow-band images generated for this process for J161306 is displayed in Figure~\ref{fig:residual}. It displays the H$\alpha$ and continuum narrowband images, along with their difference. The green pea and companion are clearly visible in the difference image. We repeated this process for the [O III] emission line as well and did not find any additional companions. We visually identified star-forming companions for 4 of the 23 galaxies. The four companions are presented in Figure~\ref{fig:spectra}, where we show their SDSS image and their MUSE spectra, alongside the green pea spectra. The plotted wavelengths are in the observed frame.

In order to calculate the error bounds on the fraction of galaxies that have companions, we had to determine what is the likelihood of the observed fraction of green peas with at least one companion. Neither a Poisson nor a Gaussian distribution would be an appropriate choice, as we present our result in the form of the number of galaxies with companions (successes) versus the number of galaxies without companions (failures), given the probability of success. In this case, the likelihood of observing 4 successes out of 23 trials follows a binomial distribution, with probability theta of single success. Using the Bayes' theorem, we can derive the posterior distribution function on theta assuming a conjugate uniform beta prior.

\subsection{MUSELET Emission Line Search}

Potential problems with the aforementioned approach are that some companions may not have strong enough emission lines to be noticed by eye. Consequently, the next step was to identify any missed emission line companions in the previous search, using a more in-depth method. This search was done using the MUSELET (MUSE Line Emission Tracker) tool from the Python package MUSE Python Data Analysis Framework (MPDAF) \citep{bacon_mpdaf_2016}. The tool uses Source Extractor \citep{bertin_sextractor_1996} and creates a number of narrow band images from the data cube, from which it can detect emission lines and separate them from continuum. It then uses multiple emission lines for a source to estimate the redshift. We applied MUSELET to the data cubes of the green peas, using the default parameters, and obtained the same four previously identified companions, but no new ones. 

\subsection{Search for companions without emission lines}

Similarly to how we searched for companions using emission lines, we performed a search for matching absorption features instead of matching emission features to find companion galaxies at a given redshift.

We used Source Extractor on the white light images to detect and identify all sources within each green pea's field of view.  We set the following values in the default.sex file to optimize the deblending: DETECT\_MINAREA: 3, DETECT\_THRESH: 2, DEBLEND\_NTHRESH: 32, DEBLEND\_MINCONT: 0.0005, SATUR\_LEVEL: 260000.0. We considered only galaxies with Source Extractor flag=2 (the object has been deblended), and that are further than 10 pixels from the edge of the field of view. We extracted the spectra of all the galaxies using an elliptical aperture with semi-major and semi-minor axis defined by Source Extractor. We set the major and minor axes of the ellipses equal to ``A\_IMAGE" and ``B\_IMAGE" parameters, multiplied by the Kron radius, which was 3.5 \citep{baronchelli_spitzer_2016}.

\begin{figure}
\plotone{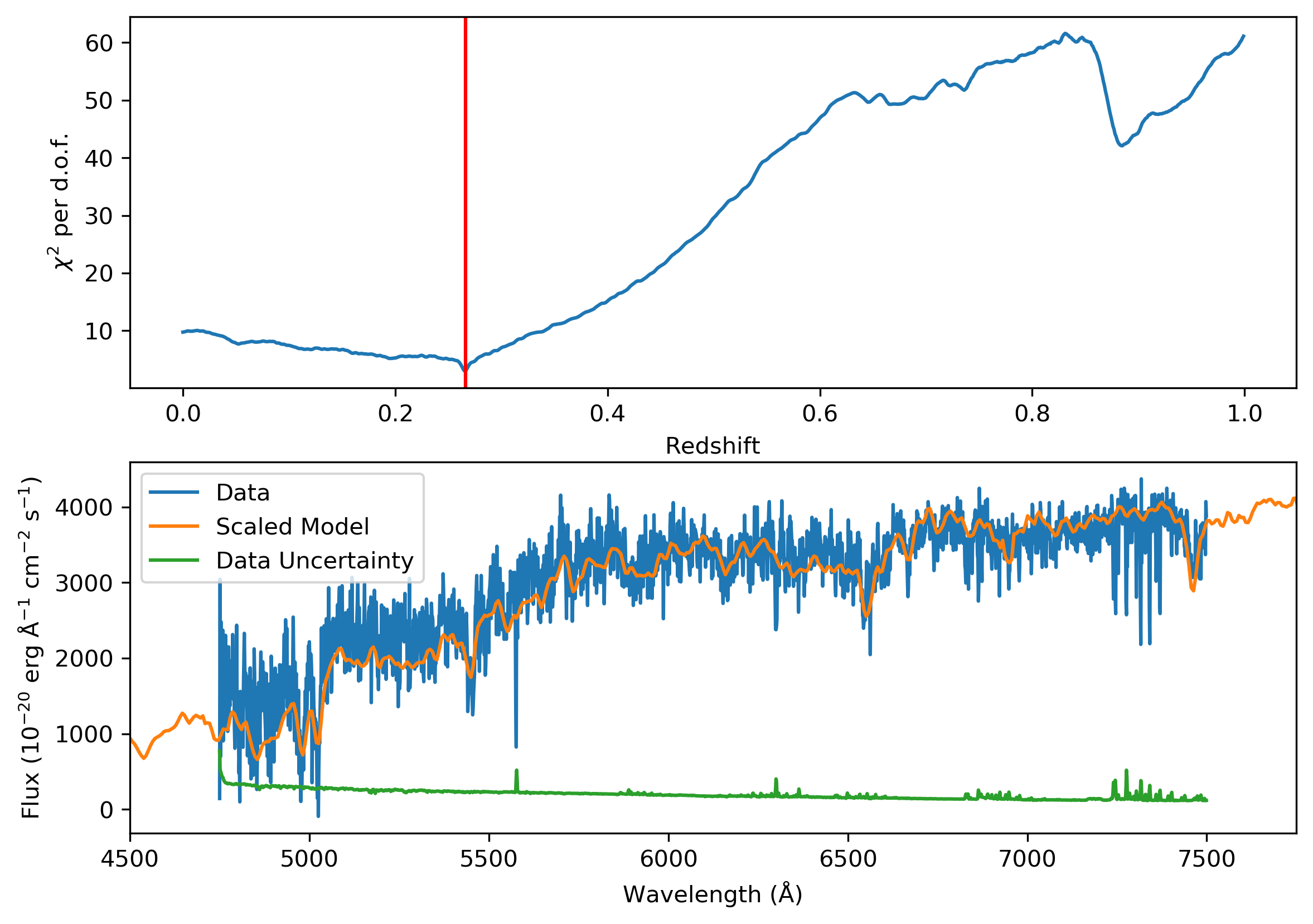}
\caption{This is an example of a background absorption galaxy put through the code to match with a template absorption spectrum and determine the redshift. The template spectrum used is the elliptical galaxy from the Kinney-Calzetti Atlas in pysynphot. The upper figure is the $\chi^2$ per degree of freedom for each redshift that the template was shifted to and compared to the object's observed spectrum. The red vertical line is the redshift where the $\chi^2$ is the lowest. The lower figure has the observed spectrum in blue, the shifted template in orange, and the uncertainty on the data in green. 
\label{fig:abs_example}}
\end{figure}

We estimated the redshift of sources without emission lines by cross-correlating the observed spectra with the elliptical galaxy
spectral template from the Kinney-Calzetti Atlas \citep{kinney_template_1996,calzetti_dust_1994}. We limited the analysis to only galaxies with an average S/N ratio in the continuum greater than 1.5. We computed the S/N in the wavelength range 5000\AA~$\le \lambda\le $~7500\AA. Figure~\ref{fig:abs_example} shows an example of a background absorption galaxy that was matched to a redshift using the absorption template matching code. The absorption line search did not result in finding any new companions. 

In summary, we identified 4 companion galaxies to 23 green pea galaxies, all identified by their prominent emission lines.

\section{Comparison to Control Sample}
\label{sec:comptocontrol}

In order to assess whether green pea galaxies have  companions more frequently than the general population of star-forming  galaxies, we build a control sample starting from the MUSE-Wide and 3D-HST survey catalogs. 

The MUSE-Wide survey \citep{herenz_muse-wide_2017, urrutia_muse-wide_2019} covers 22.2 arcmin$^2$  of the CANDELS/Deep \citep{grogin_candels_2011, koekemoer_candels_2011} area in the Chandra Deep Field South \citep{giacconi_first_2001} using the MUSE IFU spectrograph. The survey has an exposure time of one hour per arcmin$^2$ pointing, which is similar to our survey strategy. Their catalog includes more than 2,000 objects identified in this region, with either spectroscopic or photometric redshift. Relevant for this study, we consider only the  sample of galaxies with spectroscopic redshift, measured using either emission or absorption lines. 

The 3D-HST survey \citep{momcheva_3d-hst_2016,brammer_3d-hst_2012-1}, a near-infrared spectroscopic Treasury program with the Hubble Space Telescope (HST), covers $\sim600$ arcmin$^2$ using wide field grism spectroscopy (WFC3/G141 and ACS/G800L). The 3DHST catalog includes 2898 objects with spectroscopically confirmed grism redshifts in the area covered by the MUSE-Wide survey.

A number of selection effects can affect the number of companions we detect around the green peas and the comparison sample. First, we consider the depth of the data. To compare with the MUSE-Wide survey we estimated the magnitude limit of our data in the  F775W filter. We find that the 5$\sigma$ limit is $\sim$23rd magnitude, comparable with the \citet{herenz_muse-wide_2017} value. Thus the two surveys have similar depths. 

\begin{figure}
    \plotone{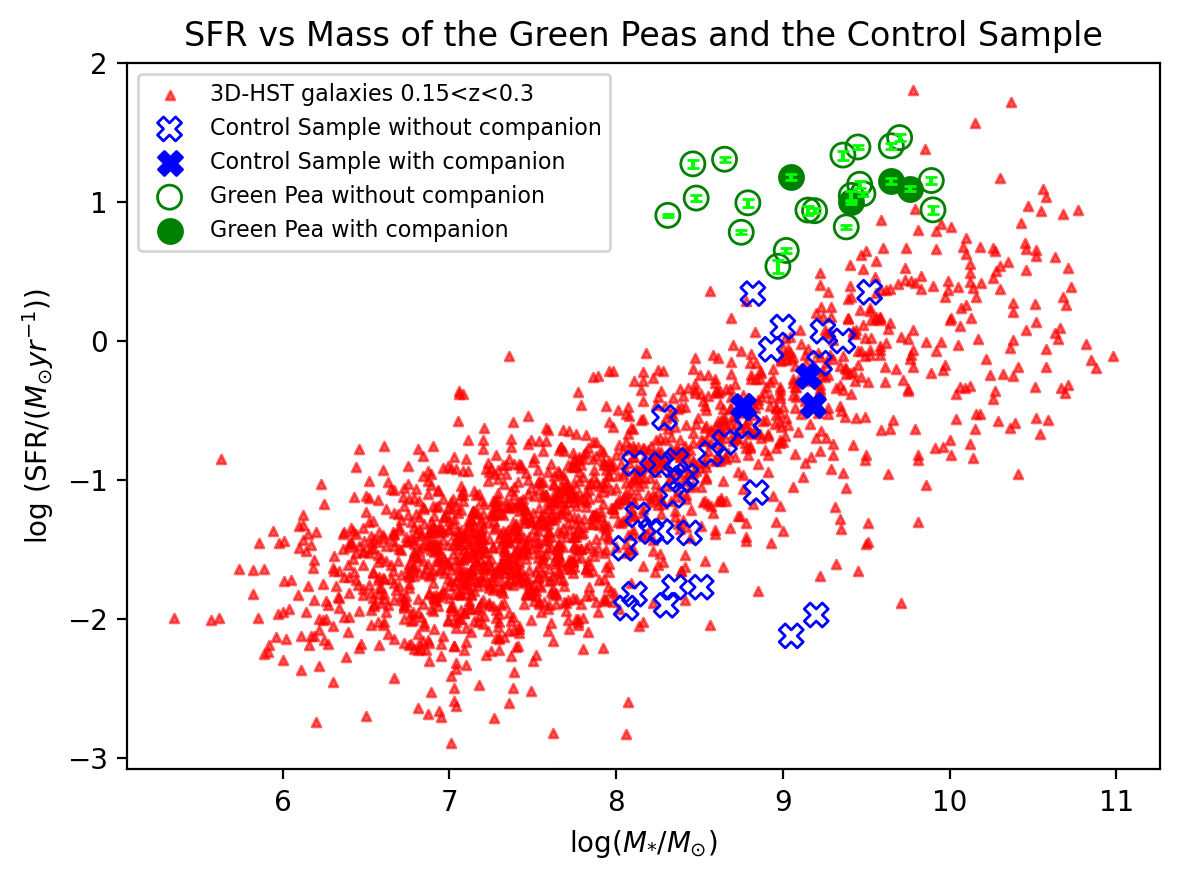}
    \caption{The star formation rates and stellar masses of the green peas in comparison to the control sample. The green peas are represented by green circles, with filled in circles for the green peas that have companions, and unfilled circles for the green peas that do not have companions. The masses used are from \citet{izotov_green_2011} when available, otherwise from \citet{cardamone_galaxy_2009} if not. The error bars on the SFR are in lime green. The blue X shapes represent the control sample galaxies from 3D-HST. The filled in X's have companions, and the unfilled X's do not have companions. The red triangles are all of the galaxies from the 3D-HST catalog that have a spectroscopically confirmed redshift between $0.15<z<0.3$. It is interesting that there are not any control sample galaxies in the mass range $>$9.5; these types of galaxies exist in the 3D-HST survey, but they are not spatially located in the MUSE-Wide footprint.}
    \label{fig:mass_sfr}
\end{figure}

We can now define the comparison sample as those galaxies in the 3D-HST survey with stellar masses and redshifts similar to those in the sample of green peas, but with star-formation rates consistent with being on  the main-sequence, for their mass. Specifically, we select those galaxies with $8 < log(M_*/M_{\odot}) < 10$, $0.15 < z < 0.3$ and SFR consistent with the bulk of the population at $z\sim 0.2$. We also require that these objects fall well within the MUSE-Wide footprint (i.e., that they are located more than 0.5 arcminutes from the survey  edges) and that their redshifts are spectroscopically measured. The SFR and stellar mass of the 43 selected objects in the comparison sample are compared with those of the green peas in Figure~\ref{fig:mass_sfr}. 

\begin{figure}
\plotone{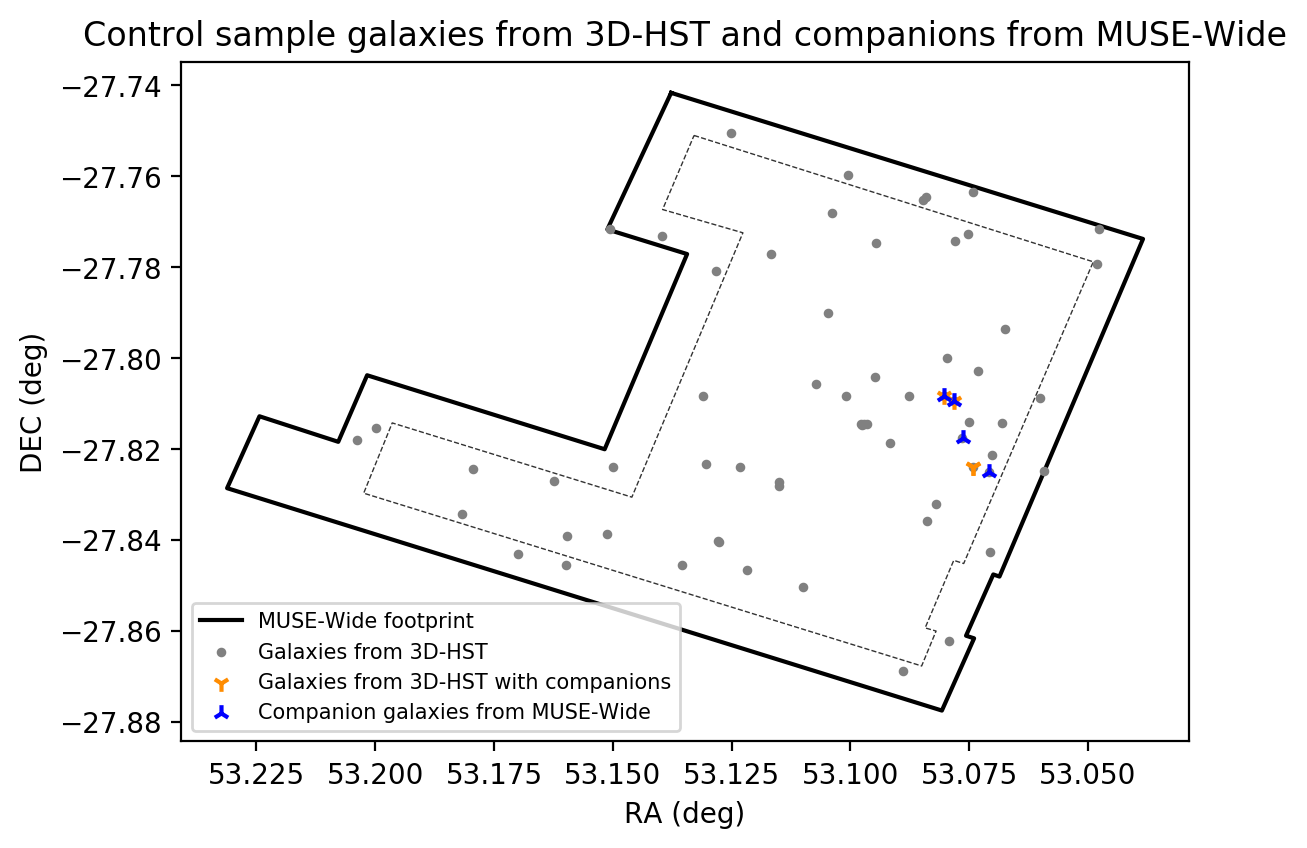}
\caption{The control sample galaxies plotted in the MUSE-Wide footprint. The grey circles are the control sample galaxies chosen from 3D-HST. Only the galaxies inside the dashed line are included, as the ones outside the dotted line could potentially have companions outside the MUSE-Wide footprint that would be missed. The orange symbols represent the control sample galaxies that have been identified as having companions, and the blue symbols represent the companion galaxies identified from MUSE-Wide. The two instances where the blue and orange overlap are actually the same pairing found twice (the pairs are not the symbols overlapping, but rather one from each). In other words, the the overlapping blue and orange symbols meant a galaxy was identified as both a galaxy in 3D-HST and a companion in MUSE-Wide. \label{fig:control}}
\end{figure}

We analyzed the comparison sample in the same way as the sample of green peas. Specifically, we searched  for the presence of any companions using the MUSE-Wide catalog, and a search radius of 0.5~arcminutes (later limited to 78~kpc, explained in Section \ref{sec:resdis}), and a velocity of $\pm300$ km s$^{-1}$. We search using the MUSE-Wide survey and not the 3D-HST data, because we want the analysis  of the companions to be comparable to our own method; the green pea companions and the control sample companions are both data taken by MUSE, and the MUSE-Wide and green pea average 3$\sigma$ flux limits are on the order of $10^{-18}$ erg~cm$^{-2}$~s$^{-1}$. We find that the control sample has a fraction of objects with companions at 3 out of 43 ($0.08_{-0.03}^{+0.05}$, 68\% credible interval calculated using a beta distribution as described in Section \ref{sec:resid}).

\section{Results and Discussion}
\label{sec:resdis}

\subsection{Is there an excess of companions to green pea galaxies?}

\begin{figure*}
\plottwo{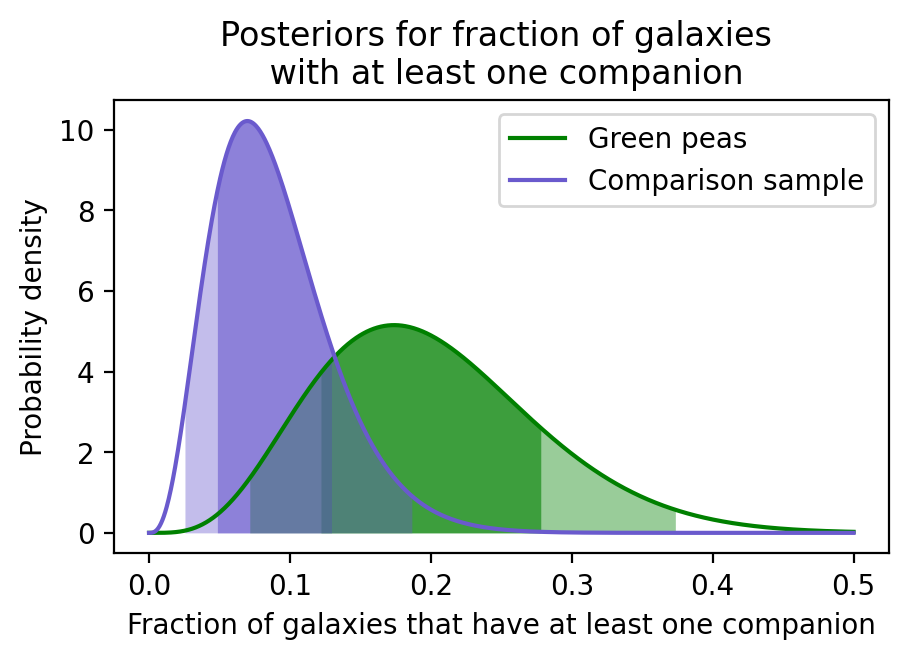}{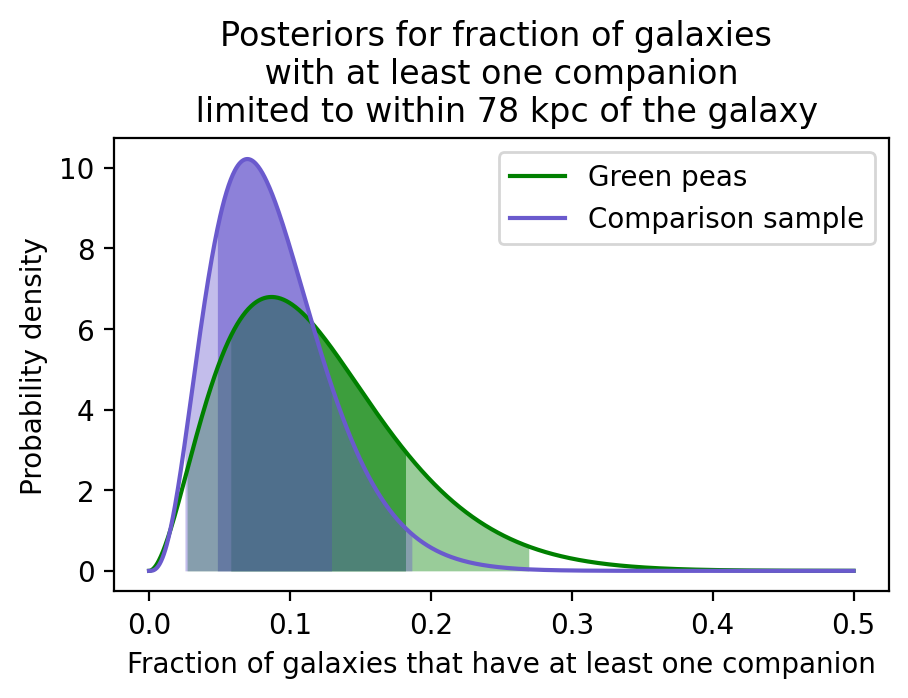}
\caption{Posterior probability distributions for the fraction of  galaxies with at least one companion. Green and purple, show green pea and control-sample, respectively. In all distributions, the dark shaded regions are the 68\% credible intervals, and the light shaded regions are the 95\% credible intervals. In the left panel we show the results computed using the full MUSE area for all galaxies, while on the right we limit the companion search to the same physical area (a circle of 78 kpc radius) at all redshifts.
\label{fig:posterior}}
\end{figure*}

By comparing the green peas with the control sample, we found that, at face value, the green peas are slightly more likely to have companions than the control sample. We find that the fraction of green peas with companions is $0.19_{-0.07}^{+0.09}$ while the fraction for the control sample with companions is $0.08_{-0.03}^{+0.05}$. The full posterior distribution functions for the fraction of galaxies with companions are plotted in Figure~\ref{fig:posterior}, left panel.

The results in Figure~\ref{fig:posterior} (left panel) are based on a search performed on the entire MUSE field of view. However, it is important to consider that the projected area differs among green peas that are at different redshifts. Half an arcminute corresponds to 78 kpc and 134 kpc at redshifts 0.15 and 0.3, respectively. The search area differs by a factor of 3 at the aforementioned redshifts. If we limit the  search to the same physical area for all of the green peas (i.e., limiting to the maximum common physical separation of 78 kpc), then we only find companions for two of the green peas, rather than four, reducing the fraction of green peas with companions to $0.11_{-0.05}^{+0.07}$. This change suggests that  we may be missing potential companions around the lowest redshift  green peas because the MUSE field of view is not wide enough. Applying the same limit to the comparison sample does not change their fraction with companions. The new posterior distribution functions on the fraction of galaxies with companions are shown in Figure~\ref{fig:posterior}, right panel. 
When the same physical scales are considered, the fraction of  green pea galaxies with companions is now indistinguishable from that computed using the control sample galaxies.

\begin{figure}[ht]
\plotone{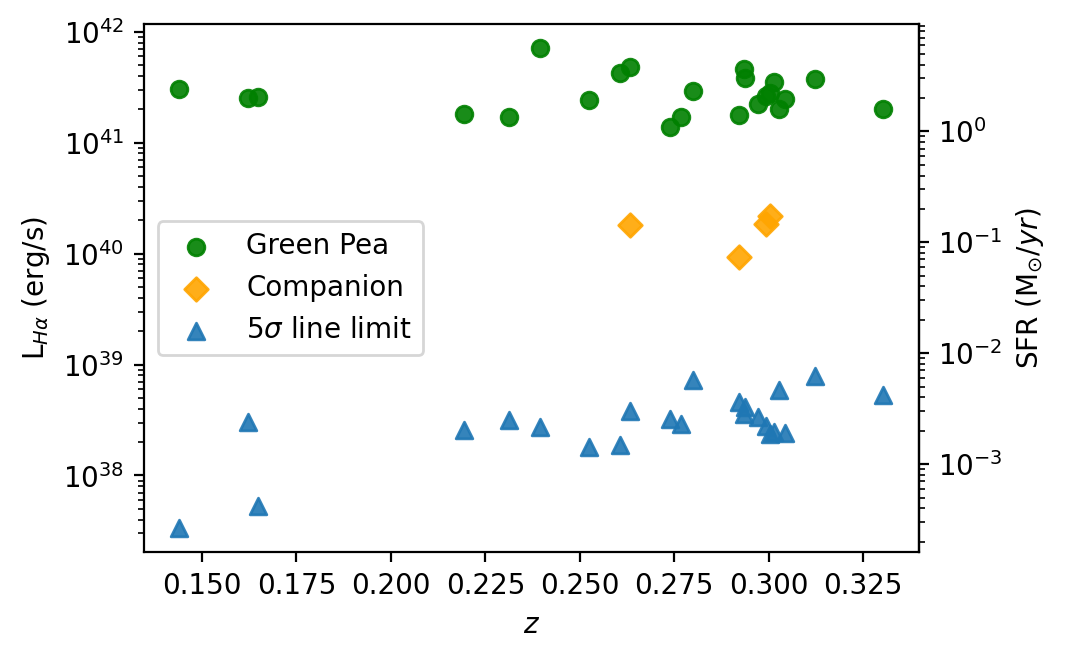}
\caption{This plot compares the H$\alpha$ luminosity of the green peas and companions with the 5$\sigma$ H$\alpha$ line luminosity limit for each green pea field. The secondary y axis is the star formation rate, assuming \cite{bell_comparison_2001}. The green peas are represented by green circles, the companions are represented by orange diamonds, and the H$\alpha$ line luminosity limits are represented by blue triangles. The luminosities of the companions are well above all the line luminosity limits. 
\label{fig:lumlim}}
\end{figure}

We also checked how the flux limits affect the detected line luminosity as a function of redshift. Indeed, if the luminosity of a companion had been lower than the limiting luminosity of a field at higher redshift, then we would need to apply constraints on the luminosity (like above with the search area). In Figure~\ref{fig:lumlim} we plot the \ha\ line luminosities for the green peas and the identified companions as a function of redshift. We also plot the 5$\sigma$ \ha\ luminosity limit for the corresponding fields. Figure~\ref{fig:lumlim} shows that all of our companions have an H$\alpha$ luminosity well above all of the line luminosity limits for our data.

An interesting point is that all of the companions are star-forming, as suggested by the presence of emission lines, indicating that they are possibly influenced by the green peas. The projected distances between the green peas and their companions, shown in Table \ref{tab:info2}, vary between 58 and 120 kpc. \citet{ellison_galaxy_2008} states that distances smaller than 30 h$^{-1}$ kpc are optimal to enhance star-formation in pairs, and  this result is in agreement with previous studies that measured the \ha~EW enhancement as a function of separation \citep{geller_infrared_2006,westra_evolution_2010,fu_sdss-iv_2018}.
Interactions can trigger starburst events because they facilitate the transport of gas toward the center \citep{shlosman_fuelling_1990, comerford_merger-driven_2015, goulding_galaxy_2018}. Using numerical simulations, \citet{noguchi_amorphous_1988} found that in the case of violent tidal interactions, clouds are driven to the nuclear region of the galaxy by the changing gravitational field associated with the changes induced by the interaction. Similarly, \citet{pan_sdss-iv_2019} notes that SFR increases with decreasing distance between two approaching galaxies, and they attribute this to the formation of non-axisymmetric structures that move the gas into the central regions of the galaxies. This enhances circumnuclear star formation. 

While 30 h$^{-1}$ kpc is smaller than the measured projected distances, it is possible that the identified companions did a flyby of the green peas, and were closer, in the past. Taking the maximum distance and maximum velocity of our identified companions, it would take about 300~Myr to travel 100~kpc at 300 km s$^{-1}$. The time-scales would imply star formation that lasted a few hundred million years. With \ha\ EW of a few hundred Angstroms (Table \ref{tab:info2}), the spectra of green peas are consistent with star formation sustained for these timescales, only under the assumption of constant star formation history. For a burst, these values in the spectra can only happen for ages younger than a few tens of Myr, under the assumption of a short duration burst (but see \citet{mcquinn_nature_2010}). The UV spectra of green peas show that O-type stars formed within the last $\sim$~5~Myr. Given that we do not find statistical evidence for an elevated companion fraction in the green peas in this study, and based upon the timescale arguments, we argue that the ``companions" are unrelated to the bursts in these galaxies.

The \hb\ EWs of the companion galaxies, as listed in Table \ref{tab:info2}, provide an interesting insight into the star formation timescales of the companions in relation to the green peas. The companions have relatively low EWs ($\lesssim 50$\AA), with the exception of the companion of J161306, where the EW(\hb) of the companion almost matches that of the green pea, which is still relatively low. These EWs are not high enough to impose significant constraints on the star formation timescales; continuous star formation over 1 Gyr cannot be ruled out by these values. These EWs are consistent with either bursts of star formation that have recently turned off \citep{leitherer_starburst99_1999}, or recently ignited bursts in an already formed stellar population; the presence of strong \oiii\ lines in the companions to J163719 and J161306 suggest there are very massive stars in place and that the star formation is young, while the absence in the companions to J094458 and J124834 suggests that the most massive stars have died and the star formation is decaying. A redshifted 4000\AA\ break is visible in the spectra of the companions to J094458 and J124834, which is consistent with a more evolved stellar population. The apparent anticorrelation between the \hb\ EW of the companion and distance between the green pea and the galaxy could indicate a previous encounter which triggered a burst that has since faded. The sample size, however, is not  large enough to draw a strong conclusion.

It is interesting to compute the number of star-forming galaxies we would have expected to identify in the searched volumes, regardless of the presence of the green pea. 
To do so, we use the H$\alpha$ luminosity function from the literature and integrate it down to the line limits of our survey. We use the Schechter luminosity  function \citep{schechter_analytic_1976}:

\begin{equation}
    \phi(L) dL = \phi^* \left(\frac{L}{L^*}\right)^{\alpha} exp\left(\frac{-L}{L^*}\right) d\left(\frac{L}{L^*}\right) 
\end{equation}

with the parameters $\alpha=-1.35^{+0.11}_{-0.13}$, $log(\phi^*(Mpc^{-3}))=-2.65^{+0.27}_{-0.38}$, and $log(L^*(erg s^{-1}))=41.94^{+0.38}_{-0.23}$ from \citet{shioya_hensuremathalpha_2008}. We integrate this function down to the the detection limits shown in Table \ref{tab:info}, and consider the volume we used for the search of companions, namely a box equal to 4.0 Mpc$^3$. We would expect to find between 0.1 -- 0.2 \ha\ emitters in the searched volume. This implies that we would expect to find between 2 to 4 out of 23 fields with at least one \ha-emitting galaxy at the considered depth. This value is consistent with the observed number of galaxies with companions.

\subsection{Comparison with previous studies}

To place our result in context, it is worth examining past studies that searched for companions near low-mass star forming galaxies. It is thought that because dwarf galaxies are too low-mass and too slowly rotating to sustain spiral arms \citep{gallagher_structure_1984}, other triggers are needed to induce the bursts of star formation that we see. \citet{taylor_h_1997} compared VLA HI observations of 21 BCD galaxies with 17 LSB dwarfs, to search for nearby companions  potentially triggering the burst of star formation in the BCDs. The BCD and LSB galaxies inhabit similar large scale environments out to 2.5 Mpc, and yet only 4/17 LSB galaxies were found to have companions compared with 12/21 BCDs within a search space of 0.25 Mpc and a radial velocity separation of 250 km s$^{-1}$. Although \citet{taylor_h_1997} did not apply a probability distribution analysis to his fraction of galaxies with companions, his results still suggest a statistically significant correlation between starbursting behavior and the presence of a companion galaxy in dwarfs. 

On the other hand, several studies have found that starbursting in dwarfs is negatively correlated with the presence of companions. \citet{telles_environment_1995} conducted an optical study of 51 BCD galaxies, comparing the morphology and brightness of those with companions to those without. They found that 12 of the 51 BCDs had a companion within 1 Mpc and 250 km s$^{-1}$, but that the galaxies with companions were of more regular morphology and lower luminosity, while those without companions tended to be brighter and have more disturbed morphology. They do acknowledge, however, that their study was conducted in the optical, where it is possible to miss low brightness companions or even just HI clouds that are acting as the interaction partner. However, \citet{telles_local_2000} performed a deeper optical search for companions, which they claimed would include such low-mass galaxies and HI clouds, and confirmed that BCDs tend to occupy low-density regions. These results are roughly in line with the result of \citet{cardamone_galaxy_2009} on green peas in particular, which found that green peas tend to be in more isolated regions of space than a comparison sample of galaxies with similar luminosity and redshift, out to 1 Mpc of projected distance.

\citet{brunker_environments_2022} studied the galaxy distribution around thirteen green peas in KISSR (KPNO International Spectroscopic Survey red) out to a distance of about 15 Mpc at the redshifts of the targets, and to a distance of 100 Mpc
on either side along the line of sight. The majority (seven) of their green peas were in extremely low density environments, which is to be expected for green peas, again in line with the result of \citet{cardamone_galaxy_2009}. They conclude that interactions and mergers are unlikely to be the main trigger of the starburst events. Their smallest nearest neighbor distance was 1.07 Mpc, which is much larger than the field of view we investigate in this paper, and they acknowledge that they are unable to resolve or rule out interactions with very nearby dwarf galaxies.

Our green peas have a companion fraction of 2/23 using a consistent distance cut at all redshifts, a much lower number than the 25\% fraction or higher found by \citet{taylor_h_1997} and \citet{telles_environment_1995}. However, our search area is constrained to a 78 kpc radius, which is substantially smaller than the 0.25 Mpc used by \citet{taylor_h_1997} and the 1 Mpc used by \citet{telles_environment_1995}. In order to accurately compare these results, a larger search area would be needed, so we cannot conclude that our green peas are more isolated than the BCDs in these studies.

The result in \citet{telles_environment_1995} is particularly interesting, as it found a negative trend between disturbed morphology and the presence of companions for dwarfs. It is possible that many of their regular BCD galaxies with a companion are in a pre-merger phase, and that interactions with their companion have not yet disturbed them. This could explain the presence of regular galaxies with companions, but cannot explain the presence of starbursting and disturbed galaxies without companions. It is fairly certain that the companion fraction of starbursting dwarf galaxies is not 100\%: even the most extreme example, \citet{taylor_h_1997}, found that only 57\% of starbursting galaxies had companions within 0.25 Mpc. We agree with the conclusion of \cite{telles_local_2000} that another explanation, aside from interactions with observable companions, is needed for the extreme star formation activity observed in green peas and starbursting dwarfs.

In the case of starbursting dwarf galaxies that don’t seem to have a visible companion, a possible trigger is mergers with dark satellites. Dark satellites are satellites that are made of dark matter and have little to no luminous component. \citet{helmi_dark_2012} ran a series of cosmological simulations and found that in the case of gas rich dwarf galaxies, encounters with dark satellites can trigger starbursts. This is worth considering as an alternate trigger for starbursting.

\section{Summary and conclusion}
\label{sec:summ}

We studied a sample of 23 green pea galaxies to identify possible faint companions that could be responsible for their strong ongoing burst of star-formation. The search was performed using  MUSE IFU data, which cover one arcminute$^2$ around each galaxy. The majority of the green peas do not have a companion within the MUSE field of view. When limiting the search to the physical area common to all galaxies (i.e., within a radius of 78 kpc), we find that the fraction of green pea galaxies with companions is $0.11_{-0.05}^{+0.07}$. We selected a comparison sample of star-forming galaxies matched to the green peas in stellar mass and redshift, and for which similar depth MUSE data were available. We found that the fraction of comparison galaxies with a companion is $0.08_{-0.03}^{+0.05}$, consistent with the value obtained for the green peas.

The companions that we identified are about two to four times further away than the expected 30 kpc distance limit for enhanced star formation \citep{ellison_galaxy_2008}. Additionally, based on timescale arguments, we conclude that it is unlikely that these companions were responsible for triggering the burst in a recent flyby. Given the lack of statistical evidence for an elevated companion fraction for the green peas compared to normal starforming galaxies, we conclude that interactions are likely unrelated to the ongoing burst of star formation in the green peas.

\begin{acknowledgments}
We thank the referee for their very useful comments and suggestions. This work was supported by the National Science Foundation Graduate Research Fellowship under Grant No. 1839286, and a Three Year Fellowship from the College of Science and Engineering at the University of Minnesota. M.H. is supported by the Knut \& Alice Wallenberg Foundation. L.L. would like to thank Dr. Vihang Mehta for his help in installing the MUSE data reduction pipeline.
\end{acknowledgments}

\software{EsoRex \citep{weilbacher_data_2020},
          MPDAF \citep{bacon_mpdaf_2016},
          Astropy \citep{the_astropy_collaboration_astropy_2018},
          Source Extractor \citep{bertin_sextractor_1996}
          }

\bibliography{GPs}{}
\bibliographystyle{aasjournal}

\end{document}